\documentclass[12pt]{iopart}

\usepackage{iopams}  
\usepackage{graphicx}
\usepackage{ulem}
\begin{document}

\title[In-situ topotactic chemical reaction for spectroscopies]{In-situ topotactic chemical reaction for spectroscopies}

\author{Tappei Kawakami$^{1,*}$, Kosuke Nakayama$^1$, Katsuaki Sugawara$^{1,2,3}$, and Takafumi Sato$^{1,3,4,5,6,*}$}

\address{$^1$Department of Physics, Graduate School of Science, Tohoku University, Sendai 980-8578, Japan.}
\address{$^2$Precursory Research for Embryonic Science and Technology (PRESTO), Japan Science and Technology Agency (JST), Tokyo 102-0076, Japan.}
\address{$^3$Advanced Institute for Materials Research (WPI-AIMR), Tohoku University, Sendai 980-8577, Japan.}
\address{$^4$Center for Science and Innovation in Spintronics (CSIS), Tohoku University, Sendai 980-8577, Japan.}
\address{$^5$International Center for Synchrotron Radiation Innovation Smart (SRIS), Tohoku University, Sendai 980-8577, Japan.}
\address{$^6$Mathematical Science Center for Co-creative Society (MathCCS), Tohoku University, Sendai 980-8578, Japan.}
\address{$^*$Authors to whom any correspondence should be addressed.}
\ead{t.kawakami@arpes.phys.tohoku.ac.jp, t-sato@arpes.phys.tohoku.ac.jp}
\vspace{10pt}
\begin{indented}
\item[]December 2023
\end{indented}

\begin{abstract}
Topotactic chemical reaction (TCR) is a chemical process that transforms one crystalline phase to another while maintaining one or more of the original structural frameworks, typically induced by the local insertion, removal, or replacement of atoms in a crystal. The utilization of TCR in atomic-layer materials and surfaces of bulk crystals leads to exotic quantum phases, as highlighted by the control of topological phases, the emergence of two-dimensional (2D) superconductivity, and the realization of 2D ferromagnetism. Advanced surface-sensitive spectroscopies such as angle-resolved photoemission spectroscopy (ARPES) and scanning tunneling microscopy (STM) are leading techniques to visualize the electronic structure of such exotic states and provide us a guide to further functionalize material properties. In this review article, we summarize the recent progress in this field, with particular emphasis on intriguing results obtained by combining spectroscopies and TCR in thin films.
\end{abstract}

%

\section{Introduction}
\subsection{Overview of topotactic chemical reaction}
 One of central research topics in modern materials science and condensed matter physics is the realization of exotic physical properties in two-dimensional (2D) or quasi-2D electron systems, as represented by the quantum Hall effect in graphene, massless Dirac transport in topological insulator (TI) surfaces \cite{NovoselovNature2005, KonigScience2007, XiaNatPhys2009, CaoNature2018, DengScience2020}, and superconductivity intertwined with spin, charge, and orbital orders in layered high-temperature ($T_{\rm c}$) superconductors and transition-metal dichalcogenides \cite{TranquadaNature1995, KamiharaJACS2008, ChuangScience2010, XiNatPhys2016}. A key to achieve such exotic physical properties partially lies on the technical advancements in obtaining atomically thin flakes and ultrathin films that utilizes, e.g., mechanical exfoliation of bulk crystals \cite{GrubiAdvSci2023}, chemical vapor deposition (CVD), and molecular-beam epitaxy (MBE) \cite{ZhangAdvSci2022, DeependraAdvMater2022}.

Recently, a sample fabrication technique that utilizes topotactic chemical reaction (TCR) is attracting a great deal of attention \cite{KageyamaNatCommun2018, CahenEJIC2019, SoodNatRevMat2021, LohNanoRes2021, LamACSNano2022, ZhangNatNanotech2023}. The TCR enables us to insert, remove, or replace atoms in a crystal while preserving one or more basic structural frameworks of the original crystal, as shown in Fig. 1 \cite{LotgeringJINC1959, ShannonNature1964, VolpeCRSE1985, FiglarzSSI1990}. Specifically, the key structural framework of the parent crystal is the square network of atoms, and it is apparently retained after inserting or replacing atoms. In the crystal after removing atoms, although the original square network is partially disrupted by vacancies, the remaining atoms form a larger square lattice that is also found in the parent crystal as a sublattice, supporting the TCR nature. The TCR is regarded as a powerful means to realize new functionalities of materials by manipulating constituent elements of the original crystals. An example of the TCR in our daily life is the lithium-ion (Li-ion) battery, which promotes reversible electronic charging and discharging associated with the movement (insertion and removal) of Li ions between the anode and cathode. Due to the TCR origin of this process, the structural framework of both the anode and cathode remains unchanged during charging and discharging, enabling the efficient storage and release of electrical energy across multiple cycles \cite{DelmasIJIM1999}. Another well-known example is perovskite transition-metal oxides, in which oxygen atoms can be exchanged with different anions such as halogen atoms by utilizing the high chemical reactivity, leading to the control of their physical properties such as magnetism \cite{KageyamaNatCommun2018}. The TCR also enables us to obtain new materials which are hardly synthesized by conventional techniques. Such capability is particularly pronounced in thin film research because the TCR typically starts at the material surface. This is exemplified by the recent successful synthesis of an infinite-layer nickel (Ni) oxide superconductor, achieved by largely removing oxygen atoms from the non-superconducting (Nd,Sr)NiO$_3$ thin films through the TCR \cite{LiNature2019}. An additional great advantage of the TCR is its low reaction temperature which is helpful to fabricate high-quality semiconductor heterostructures and plays an important role in the industrial applications of next-generation semiconductor devices \cite{ZhangNatNanotech2023}.

\begin{figure}
\begin{center}
\includegraphics[width=3.9in]{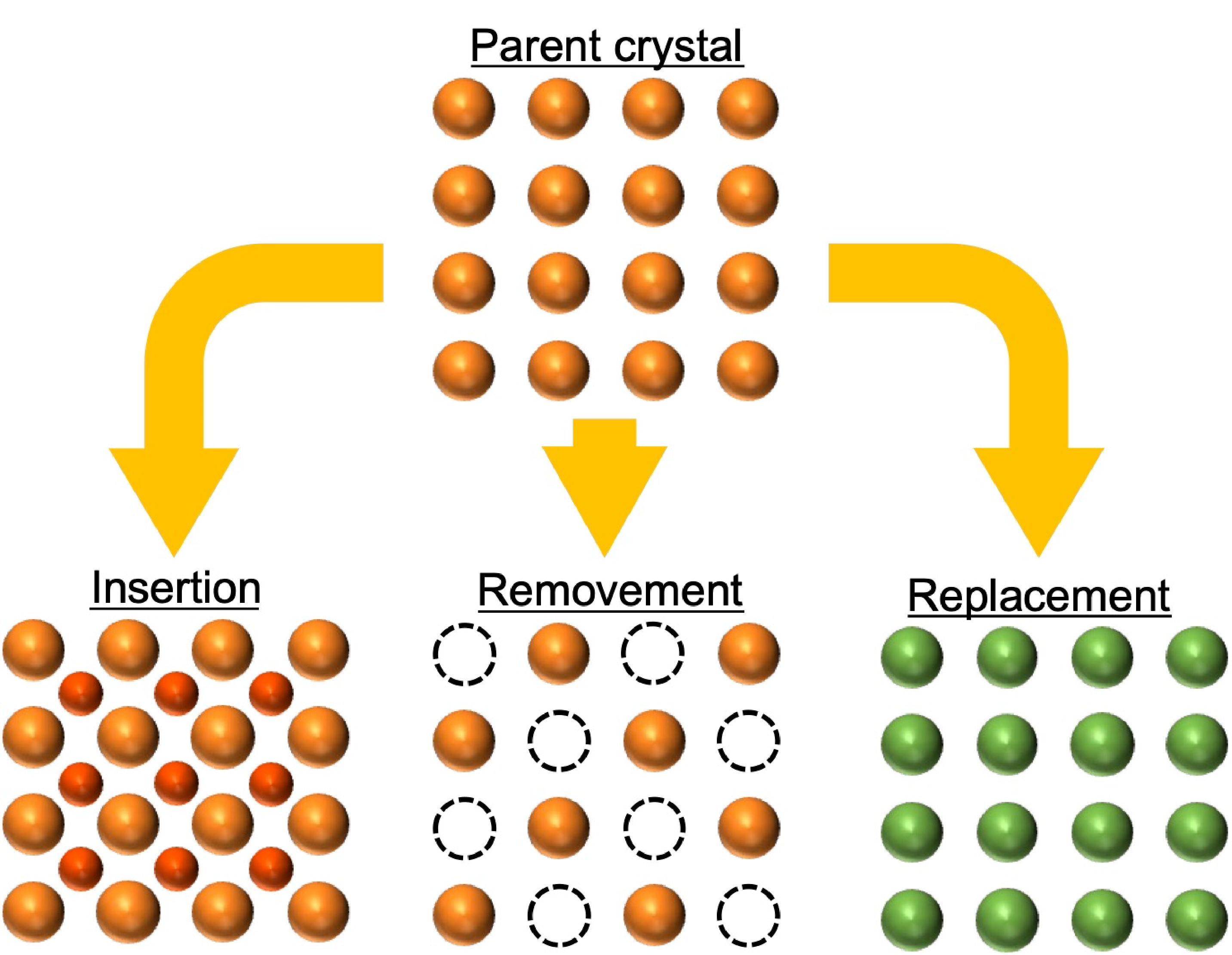}
\caption{Schematics of TCR that involves (left) insertion, (middle) removal, and (right) replacement of atoms.}
 \end{center}
\end{figure}

An important issue in utilizing the TCR is how one can promote the TCR while maintaining good crystallinity of the original sample. In this respect, key ingredients for successful TCR include (i) parent materials, (ii) reactants, and (iii) reaction environment. Regarding (i), thin films and surfaces of bulk crystals are usually selected as a platform of TCR because, as mentioned above, the reaction starts from the surface so that a high surface-to-volume ratio of parent materials is well suited for pronounced TCR. Regarding (ii), in the TCR of insertion process, atoms with small radii are generally suitable as reactants because they can be easily incorporated into a target crystal. In addition, electron-donating nature of reactants is suggested to promote their insertion into layered materials \cite{CahenEJIC2019}. Therefore, alkali metals with small atomic numbers are often used as reactants, as in the case of the Li-ion battery. Notably, while alkali-earth metals with larger atomic radius are usually more difficult to be inserted into the layered materials, this problem can be solved by the multi-step TCR. Namely, first, alkali metals are inserted into the layered materials, and then they are fully replaced with the alkali-earth metals \cite{KanetaniPNAS2012}. For the replacement process, atoms with a large electronegativity (such as halogens) are often used as reactants, in particular, in an ionic crystal. This is because a large difference in the electronegativity between the reactants and the anions in the parent crystal can promote their exchange \cite{Kitteltext}. As for the atom removal, strong reducing and oxidization agents are utilized. The removal of atoms can be also facilitated by annealing the crystal. Regarding (iii), the TCR can be ideally performed in an environment where only the parent materials and reactants exist. Therefore, inert gases or ultrahigh vacuum (UHV) conditions are preferred to avoid contamination of the sample during the TCR. Moreover, the TCR is efficiently prompted by precisely controlling the temperature of the parent material and reaction rates of the reactants.

Since thin films and surfaces of bulk materials are a primary research focus utilizing the TCR, the characterization of these materials often necessitates surface sensitive experimental techniques. Recently, the TCR has been combined with the surface sensitive spectroscopies such as angle-resolved photoemission spectroscopy (ARPES) and scanning tunneling microscopy (STM) to explore exotic quantum states and new crystalline phases. This article reviews recent progress in the rapidly growing field that combines the {\it in-situ} spectroscopies and the TCR-utilized sample growth techniques.

\subsection{Combination with spectroscopies}

\begin{figure}
\begin{center}
\includegraphics[width=3.9in]{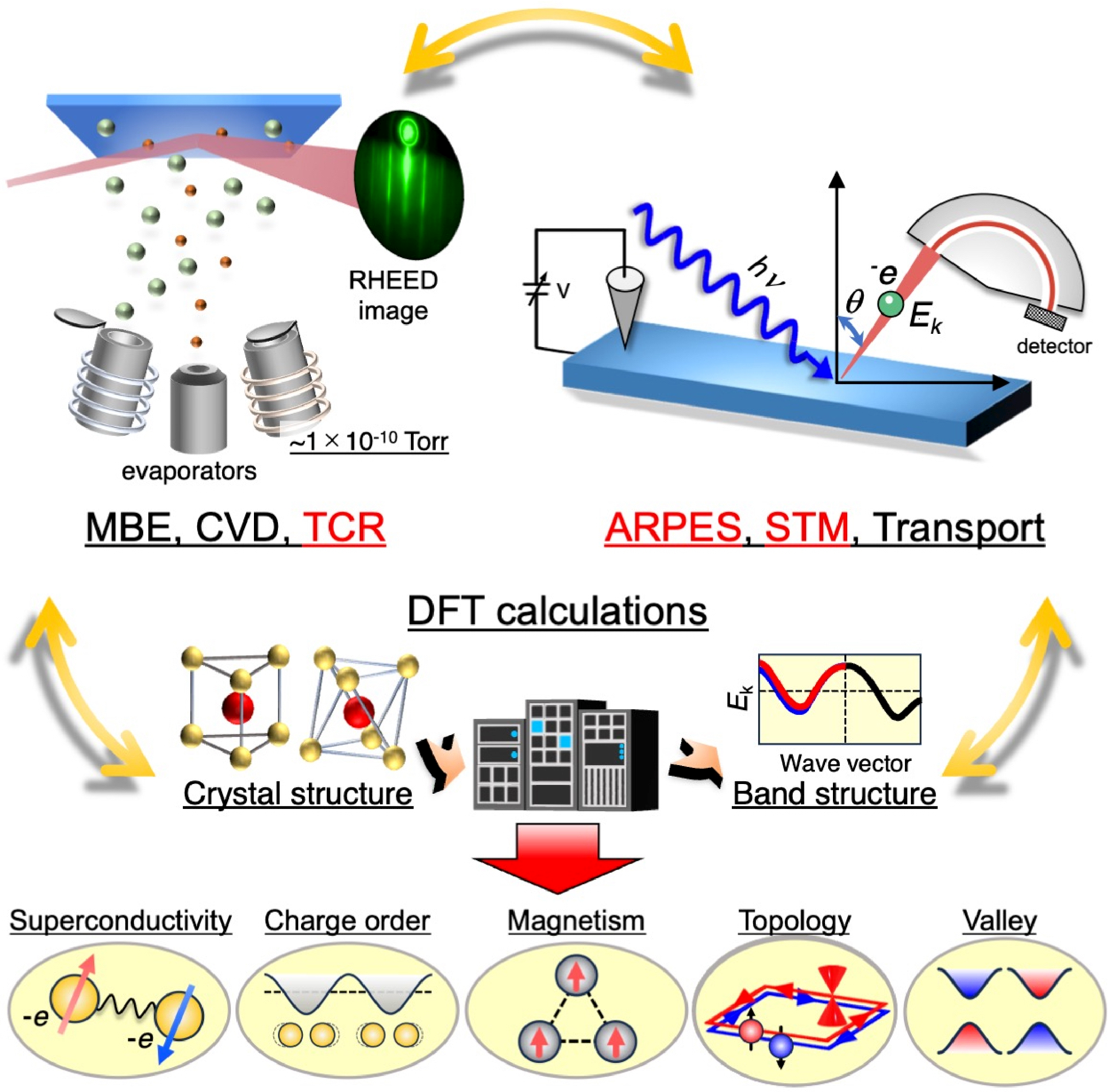}
\caption{Schematics to combine sample fabrication (MBE, CVD, TCR), characterization (ARPES, STM, transport), and DFT calculations. Integration and iteration of these key techniques accelerates exploration of new quantum materials that host intriguing physical properties such as superconductivity, charge order, magnetism, topological properties, and valley feature.}
 \end{center}
\end{figure}

Spectroscopy techniques such as ARPES and STM serve as indispensable experimental tools for directly elucidating the electronic structure of solids which is responsible for their physical properties. By comparing experimental and DFT (density-functional-theory)-derived band structures, one can deepen the understanding of underlying mechanisms of physical properties. When spectroscopies and DFT calculations are further integrated into sample fabrication techniques with tunable composition and structural parameters, they become even more powerful for exploring exotic quantum properties, as exemplified by the discoveries of unconventional superconductivity, 2D magnetism, non-trivial topological phases, and spin-valley-locked states (Fig. 2) \cite{GrubiAdvSci2023, DeependraAdvMater2022,  NakataNatCommun2021}.

Uncovering the fine electronic structure through ARPES and STM requires the preparation of a high-quality single crystal with an atomically flat clean surface. Therefore, the sample surfaces are typically prepared and treated under UHV conditions. One can satisfy this requirement through several techniques such as the MBE growth, ion sputtering, and cleaving single crystals under UHV \cite{ZhangNRMP2022}. Besides these techniques, the utilization of TCR for preparing clean surfaces and thin films enriches the category of accessible materials under UHV. For example, while the MBE method typically has difficulty in evaporating elements with high vapor pressure, the TCR enables to introduce such elements into the crystal through the replacement while maintaining good crystallinity of the parent crystal, thereby making it possible to study materials that can hardly be grown by the MBE method \cite{ShigekawaPNAS2019}. In atom-inserted layered materials, the presence of 3D covalent bonds sometimes makes it difficult to prepare flat surfaces by the cleaving. On the other hand, the TCR sometimes makes the surface preparation easier, because it enables the insertion of atoms between the layers in UHV. Thus, the TCR is not only suitable to fabricate materials which are hard to be accessed by the conventional techniques, but also becomes powerful when it is combined with the {\it in-situ} spectroscopy techniques that visualize key electronic states responsible for the exotic physical properties \cite{KanetaniPNAS2012, ShigekawaPNAS2019, HiraharaNatCommun2020, BaoPRL2021, KusunoseAPL2022, FujisawaAdvMater2023, Kawakaminpj2DMA2023}.

\subsection{Application to quantum materials}
From this section, we show actual examples of {\it in-situ} spectroscopy studies combined with the sample preparation utilizing the TCR. As a precursor of TCR, layered materials such as graphite (as well as graphene), transition-metal dichalcogenides (TMDCs), topological insulators (TIs), and Fe-based superconductors (FeSCs) are often chosen. This is mainly because of the existence of a van der Waals gap between the layers which promotes the TCR with the insertion of alkali-metal atoms and the removal/replacement of anion atoms. Also, in addition to the bulk crystal, thin films with homogeneous surface and precisely controlled thickness can be obtained in these systems, effectively enhancing the performance of TCR. Figure 3 highlights the crystal structure after TCR (insertion, removal, and replacement of atoms) on representative quantum materials, i.e. graphene, TMDCs, TIs, and FeSCs.

\begin{figure}
\begin{center}
\includegraphics[width=3.9in]{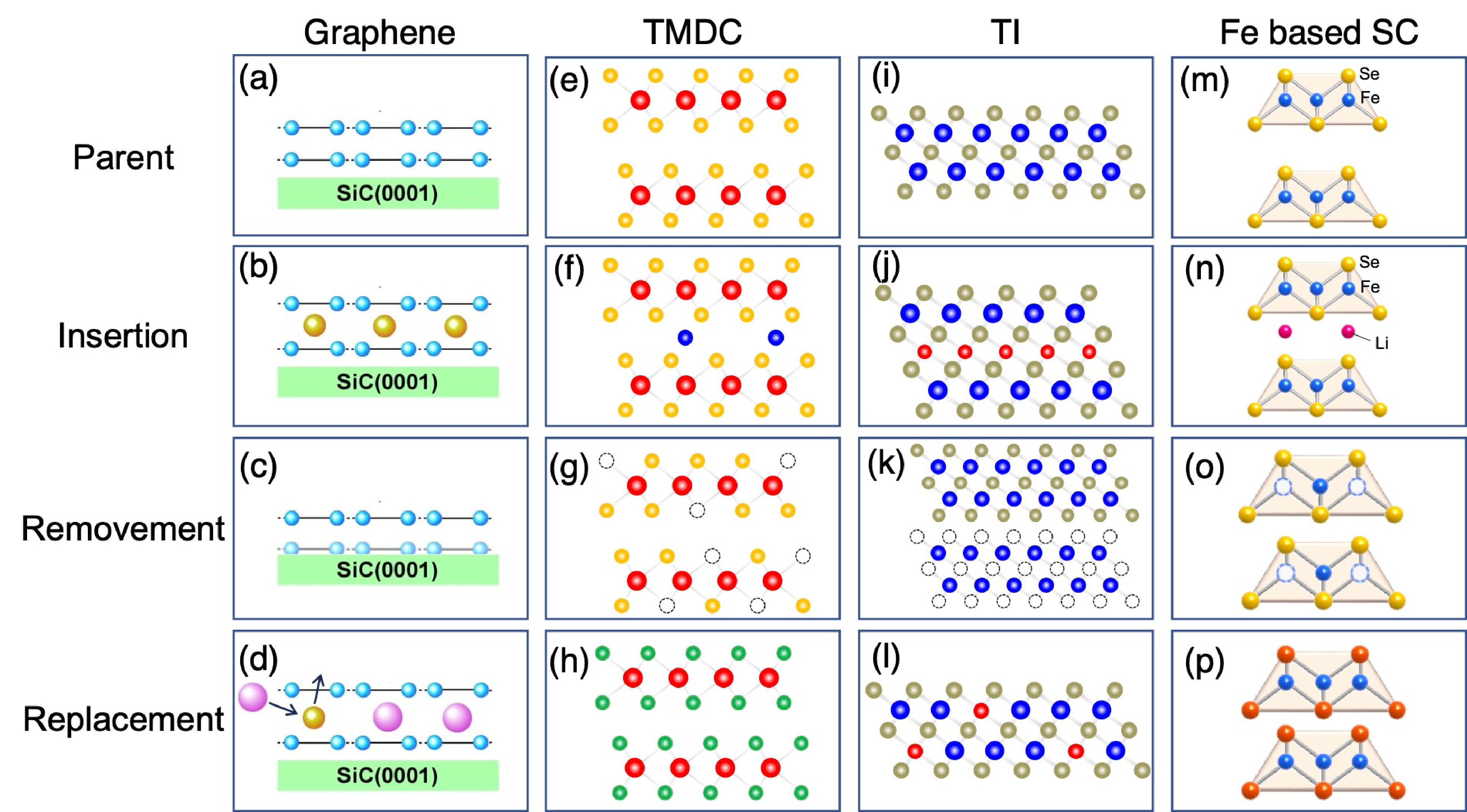}
\caption{(a)-(d) Schematics to show how crystal structure change from (a) bilayer graphene on SiC(0001) (parent material) upon (b) insertion and (c) removal. (d) Replacement of pre-inserted atoms in (b) with different atoms. (e)-(p) Same as (a)-(d) but for (e)-(h) TMDCs, (i)-(l) TIs, and (m)-(p) and FeSCs.}
 \end{center}
\end{figure}

In bilayer graphene grown on SiC(0001) [Fig. 3(a)], various elements such as Li atoms can be inserted between the graphene layers through TCR to realize bilayer-graphene intercalation compounds and novel 2D sheets encapsulated by graphene [Fig. 3(b)] \cite{SugawaraAIPAdv2011, YagyuAPL2014, AndersonRPM2017, YajiPRL2019, BriggsNatMater2020}. Also, lighter and smaller hydrogen (H) atoms can be inserted between the SiC substrate and the graphene sheet. In this case, buffer C layer which is strongly coupled to the substrate becomes free-standing and single-layer graphene on SiC is converted to free-standing bilayer graphene on H-terminated SiC. Reversely, when H atoms are removed by post annealing, bilayer graphene turns again into monolayer graphene \cite{RiedlPRL2009} [Fig. 3(c)]. It is also possible to replace pre-inserted alkali-metal atoms with other atoms having larger atomic radius, as represented by calcium (Ca) inserted bilayer graphene (C$_6$CaC$_6$) showing 2D superconductivity [Fig. 3(d)] \cite{KanetaniPNAS2012}. In TMDCs [Fig. 3(e)], different types of atoms can be inserted into the van der Waals gap [Fig. 3(f)] \cite{BronoldAPA1991, BrauerRPB1997, NakataPRM2019} as in the case of graphene. Also, by partially removing chalcogen atoms from TMDCs [Fig. 3(g)], transition-metal atoms can be self-inserted into the van der Waals gap, leading to rich electronic phases \cite{BonillaAMI2020, LasekAPR2022}. In TMDCs, the existence of ionic bonding between transition metal and chalcogen atoms enables us to replace elements with other anions having higher electronegativity, such as sulfur (S) and selenium (Se) atoms [Fig. 3(h)] \cite{Kawakaminpj2DMA2023, Xia2DMater2018, BarjaNatCommun2019, ChenACSNano2019}. This enables us to explore exotic quantum phases. In TIs with layered structure such as Bi$_2$Se$_3$ [Fig. 3(i)], guest atoms like Cu and Sr can be inserted within the van der Waals gap to turn the ground state into a superconductor \cite{HorPRL2010}. In addition, magnetic layers can be incorporated into Bi$_2$Se$_3$ to realize magnetic TI [Fig. 3(j)] \cite{HiraharaNanoLett2017}. One can also remove chalcogen layers to realize a superlattice containing TI layer and Bi bilayer [Fig. 3(k)] \cite{KusakaAPL2022}. Replacement of elements (group-V elements or chalcogen atoms) can be made in TIs to control the strength of spin-orbit coupling which is linked to their topological properties [Fig. 3(l)]. In FeSCs [Fig. 3(m)], superconducting properties can be highly modulated by the alkali-metal insertion [Fig. 3(n)] \cite{PhanJPS2017}. By partially removing Fe atoms from FeSe, a new superstructure associated with the vacancy order of Fe atoms can be realized [Fig. 3(o)] \cite{FangPRB2016}. As in the case of atomic-layer TMDCs, the Te atoms can be replaced with the S or Se atoms to convert FeTe to FeS or Fe(Se,Te) [Fig. 3(p)], and they serve as a  test to pin down the mechanism of superconductivity \cite{ShigekawaPNAS2019, WeiNanoRes2023}. As represented by these examples, TCR combined with spectroscopies has been successfully applied to investigate the exotic quantum states. In the following, we provide overview of the key experiments carried out thus far by categorizing the results by types of TCR, i.e. insertion (section 2), removal (section 3), and replacement of atoms (section 4).

\section{Insertion of atoms}
\subsection{Intercalated graphene compounds}
  Graphite is a well-known system in which various atoms and molecules can be inserted between adjacent graphene sheets through TCR. The resulting  products are known as graphite intercalation compounds (GICs) and have been studied for several decades. GICs exhibit various physical properties that depend on inserted atoms and molecules, as represented by the realization of superconductivity, magnetism, and charge-density wave (CDW). Besides showing intriguing physical properties, GICs have industrial applications such as Li ion batteries \cite{EbertARM1976, DresselhausAdvPhys1981}. Initiated by the discovery of Dirac fermions in monolayer graphene \cite{NovoselovNature2005}, numerous attempts have been made to realize graphene intercalation compounds through inserting various atoms, such as Li, K, Ca, Fe, Sn, Pb, and Bi, into bilayer graphene and monolayer graphene on SiC \cite{VirojanadaraPRB2010, LudbrookPNAS2015, HuempfnerAMI2022, ToyamaACSNano2022, McChesneyPRL2010, WarmuthPRB2016, KShenJPCC2018}. As a result, Dirac-fermion properties have been modified by the insertion. Here we show spectroscopic studies that focus on the insertion into a few-layer graphene grown on SiC, which hosts $\pi$/$\pi^*$ band forming a Dirac-cone-like energy dispersion at the Brillouin-zone corner (K point) [Fig. 4(a)]. Sugawara {\it et al.} have succeeded in fabricating Li-intercalated bilayer graphene (C$_6$LiC$_6$) by {\it in-situ} deposition of Li atoms onto bilayer graphene on SiC at room temperature under UHV. This is evidenced by the observation of folded $\pi$/$\pi^*$ bands at the zone center ($\Gamma$ point) that follow the periodicity of ordered Li intercalants, as shown in Fig. 4(b) \cite{SugawaraAIPAdv2011}. They also found a downward energy shift of the Dirac-cone band, demonstrating effective control of electron carrier density through Li insertion. In C$_6$LiC$_6$, Bao {\it et al.} further discovered the gap opening at the Dirac point due to the chiral-symmetry breaking associated with the formation of a Kekul\'{e}-ordered structure and the coupling of intervalley electrons [Figs. 4(c)-(e)] \cite{BaoPRL2021}. Besides the Li case, Ca- and K-intercalated graphene were also fabricated, and they exhibited a flat-band-like structure at the M point associated with the van-Hove singularity (i.e. saddle point) of the $\pi^*$ band [Fig. 4(f)] due to a higher electron carrier density than the Li-intercalated case \cite{McChesneyPRL2010}. This system is interesting because chiral $d$-wave superconductivity was theoretically predicted to show up when the chemical potential is situated at the van-Hove singularity \cite{NandkishoreNatPhys2012}. In this context, heavily electron doped nature of Ca- and K-intercalated graphene would be more suited compared to the lightly-doped Li case to realize possible unconventional superconductivity. 
 
 \begin{figure}
\begin{center}
\includegraphics[width=3.9in]{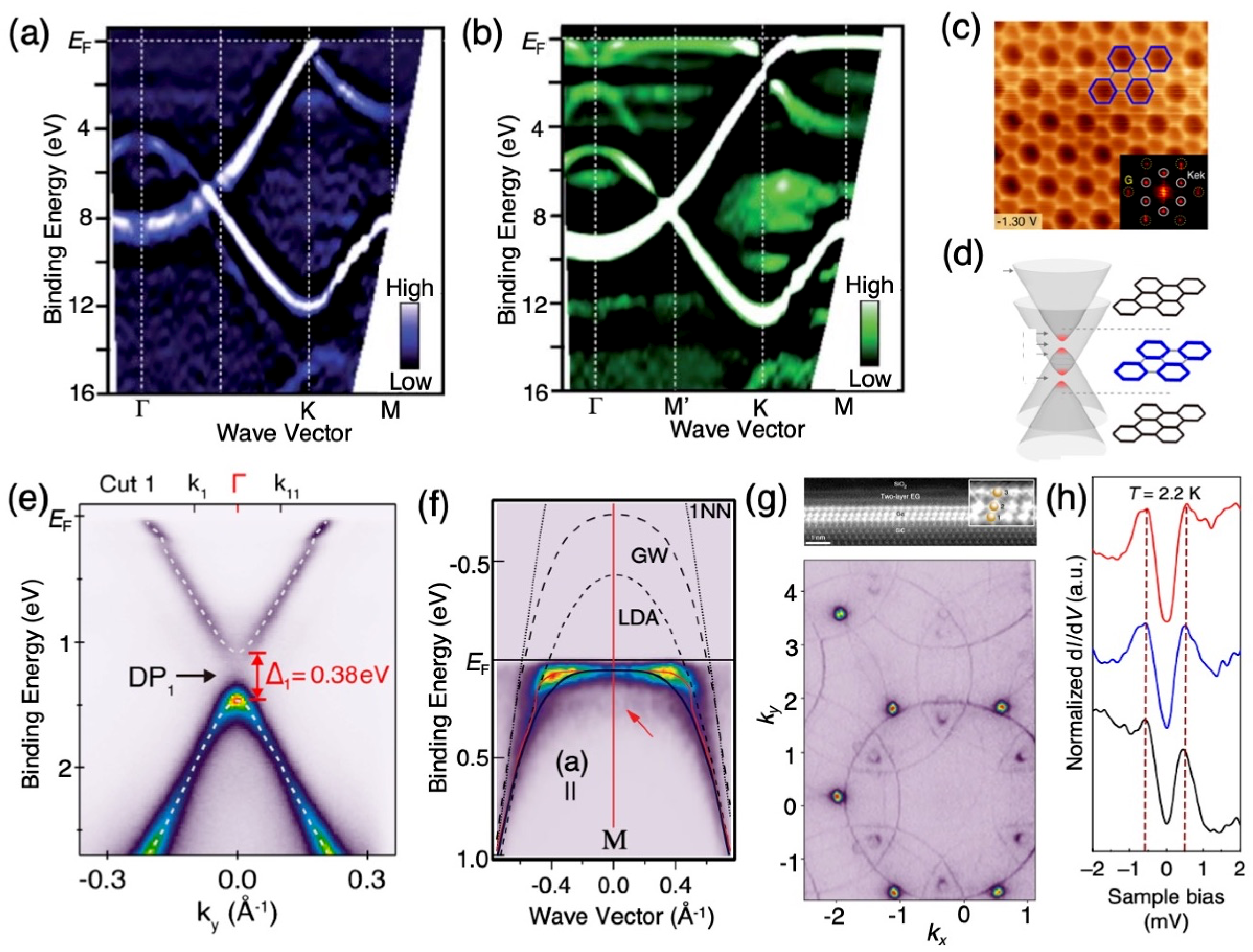}
\caption{(a), (b) Experimental band structure of pristine and Li-inserted bilayer graphene (C$_6$LiC$_6$), respectively, obtained by taking the second derivative of ARPES intensity as a function of wave vector and binding energy \cite{SugawaraAIPAdv2011}. (c) STM topographic images near the binding energy of Dirac-cone state. (d) Schematics of Kekul\'{e}-O type order. (e) Schematics of folded Dirac cones measured along the M$\Gamma$M cut \cite{BaoPRL2021}. (f) ARPES intensity of K- and Ca-dosed graphene along the KMK' cut compared with the calculated energy bands \cite{McChesneyPRL2010}. (g) Top: cross-sectional STEM image for three layers of Ga inserted between epitaxial graphene and SiC substrate. Bottom: ARPES intensity at $E_{\rm F}$ plotted as a function of 2D wave vectors. (h) Differential conductance ($dI$/$dV$) spectra for Ga-inserted graphene/SiC \cite{BriggsNatMater2020}.}
 \end{center}
\end{figure}

Recently, multilayers of gallium (Ga) and indium (In) were shown to be inserted between graphene and SiC [Figs. 4(g) and 4(h)] \cite{BriggsNatMater2020}. This was enabled by directly annealing Ga or In granules and graphene/SiC together under Ar-gas atmosphere. Since atomic-layer Ga and In show superconductivity at low temperatures, they serve as an excellent platform to study the superconducting proximity effect to the electronic structure of graphene, as represented by the observation of a pairing gap by STM in Fig. 4(h). It is also possible to utilize this graphene-Ga-SiC system as a template to fabricate a TI thin film on top of it to study possible topological superconductivity that utilizes superconducting proximity effect on the topological Dirac-cone states \cite{LiNatMater2023}.

\subsection{Magnetic TIs}
Insertion of various atoms into the van der Waals gap of TIs has been intensively performed thus far (see e.g. \cite{KoskiJACS2012, ChaNanoLett2013, YaoNatCommun2014}) to realize functional properties such as topological superconductivity, as exemplified by Cu-doped Bi$_2$Se$_3$ \cite{HorPRL2010}. Evolution of electronic states upon insertion was intensively studied by electron spectroscopies \cite{WrayNatPhys2010, TanakaPRB2012}. In the ultrathin films, {\it in-situ} insertion of Cu atoms into TIs is known to be rather difficult compared to the bulk case, whereas the insertion of magnetic layers was reported to be possible. It is known in the TIs that realization of spontaneous ferromagnetism by introducing magnetic atoms into the crystal is a key ingredient to realize exotic topological phenomena such as the quantum anomalous Hall effect (QAHE) which was shown to be achieved by doping magnetic ions into the crystal. Since the QAHE can be realized by situating the chemical potential within an energy gap of the Dirac-cone surface state (Dirac gap) associated with the time-reversal-symmetry breaking, significant efforts have been devoted to visualize the Dirac gap by spectroscopy techniques, e.g., in Cr-doped (Bi,Sb)$_2$Te$_3$ \cite{ChangScience2013}. On the other hand, the magnetic impurity doping has an inherent problem because it would cause a degradation of the sample quality, limiting the observation temperature of the QAHE to be at most a few Kelvin. An approach to potentially overcome this problem may be insertion of magnetic layers into the crystal, which is called ``magnetic extension'' \cite{Otrokov2DMater2017, OtrokovJEPTL2017} where TCR was turned out to play a crucial role. Hirahara {\it et al.} applied this technique to the thin film of prototypical tetradymite TI, Bi$_2$Se$_3$ \cite{HiraharaNanoLett2017} which consists of a stacked Se-Bi-Se-Bi-Se quintuple layers (QL). They have succeeded in inserting a Mn layer by depositing Mn atoms on top of a Bi$_2$Se$_3$ film in the Se atmosphere with annealing the sample at $\sim$240 $^{\circ}$C to fabricate a MnBi$_2$Se$_4$/Bi$_2$Se$_3$ heterostructure in which the topmost layer has a sequence of Se-Bi-Se-Mn-Se-Bi-Se [Fig. 5(b)]. Unlike pristine Bi$_2$Se$_3$ with a gapless Dirac-cone surface state, they identified a Dirac gap of $\sim$100 meV [Fig. 5(a)] which is reproduced by the DFT calculation [Fig. 5(b)]. This method was also applied to another isostructural TI, Bi$_2$Te$_3$, leading to the realization of MnBi$_2$Te$_4$ and Mn$_3$Bi$_2$Te$_7$ layers on top of the Bi$_2$Te$_3$ film and the observation of the Dirac gap persisting up to $\sim$200 K \cite{HiraharaNatCommun2020}.

\begin{figure}
\begin{center}
\includegraphics[width=3.9in]{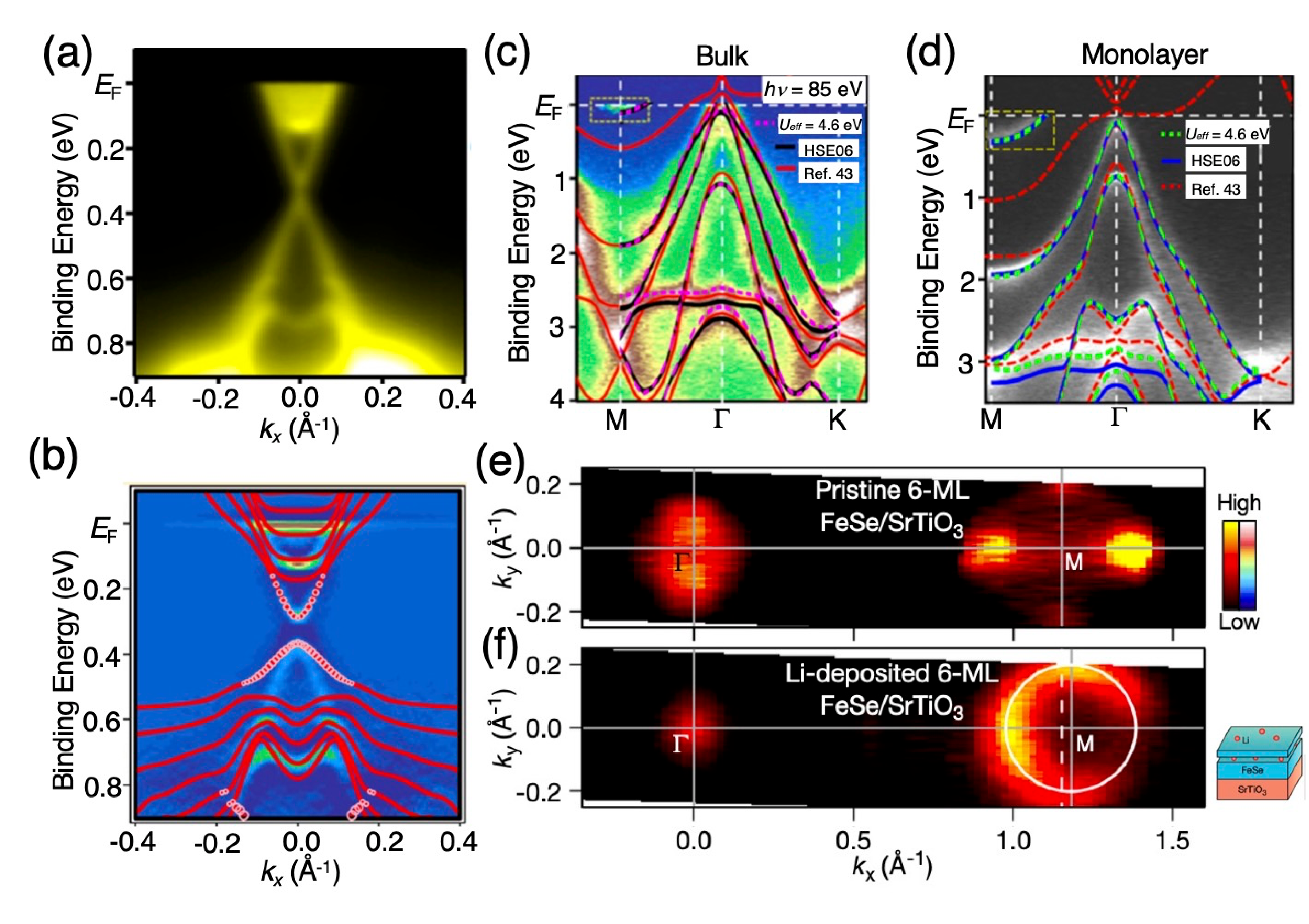}
\caption{(a) Band dispersion near $E_{\rm F}$ of MnBi$_2$Se$_4$-terminated Bi$_2$Se$_3$ thin film measured along the $\Gamma$M cut at $h\nu$ = 21 eV. (b) The calculated band dispersion overlapped with the experimental data \cite{HiraharaNanoLett2017}. (c), (d) ARPES intensity plots of bulk and K-inserted 1T-HfTe$_2$ together with the calculated band structure obtained with the DFT calculation using different approximations (GGA, HSE06, and GGA+$U$ with an effective Hubbard parameter $U_{\rm eff}$ = 4.6 eV) \cite{CruzadoACSAEM2021}. (e), (f) ARPES-intensity mapping at $T$ = 30 K for as-grown and Li-deposited 6 monolayer FeSe film plotted as a function of 2D wave vectors \cite{GiaoJPSJ2017}.}
 \end{center}
\end{figure}

\subsection{Atom-inserted TMDCs}
  The TMDCs serve as an excellent platform to insert various atoms due to the layered structure and the presence of van der Waals gap between the unit MX$_2$ (M: transition metal, X: chalcogen) layers. The insertion leads to the unique structural phases and characteristic physical properties district from those of the parent compounds. Indeed, various intercalated bulk TMDC compounds exhibit exotic quantum states, and the dimensionality (3D or 2D) of electronic structure can be also altered by the insertion of atoms \cite{BronoldAPA1991, NakataPRM2019, ZhangNatPhys2022}. So far, insertion of alkali metals and transition metals through TCR was intensively carried out in the TMDCs \cite{ZhangNatCommun2018}. As a representative, we show evolution of the band structure for bulk 1T-HfTe$_2$ upon K insertion \cite{NakataPRM2019}. Bulk 1T-HfTe$_2$ [Fig. 5(c)] exhibits hole and electron bands crossing $E_{\rm F}$ at the $\Gamma$ and M points, respectively, due to its semi-metallic nature. In contrast, K-inserted 1T-HfTe$_2$ [Fig. 5(d)] displays a significant energy shift in these bands due to the electron doping from the K atoms. The K insertion also leads to the significant change in the band structure at higher binding energies. Intriguingly, the experimental valence-band structure shows a good matching with the DFT-derived band structure for monolayer 1T-HfTe$_2$ [red dashed curves in Fig. 5(d)]. This indicates the change in the dimensionality from 3D to 2D upon K insertion. It is noted that the semimetallic band overlap is overestimated in the GGA calculation \cite{NakataPRM2019}, as can be seen from the apparent mismatch on the bottom of electron pocket between the experiment and calculation (red curve) at the M point. This problem was solved by adopting the HSE06 functional in the DFT calculations (blue curves) \cite{CruzadoACSAEM2021}.

\subsection{Atomic-layer FeSCs}
Iron selenide (FeSe) is the structurally simplest FeSC, consisting solely of conduction layers, and exhibits a relatively low $T_{\rm c}$ of 9 K \cite{HsuPNAS2008}. While a common approach to explore high $T_{\rm c}$ in FeSCs involves tuning carrier concentrations, achieving such tuning in FeSe was initially challenging due to the absence of a charge-providing or reservoir layer in the crystal. However, it turned out that the insertion of alkali metals or molecules between the FeSe layers is an effective method of electron doping \cite{GuoPRB2010, WangPRB2011, FangEPL2011, YingSciRep2012, ScheidtEPJB2012, KrztonJPCM2012, BurrardNatMater2013, NojiPhysC2014}, and for this purpose, the TCR has been found to be helpful. For example, the insertion of alkali metals can be accomplished by a soft chemical method, such as soaking bulk FeSe crystals in liquid ammonia containing alkali metals, or an electrochemical method by applying voltage in an electrolyte with alkali metal as the anode and FeSe crystal as the cathode. Additionally, Phan {\it et al.} demonstrated Li insertion by simply depositing Li atoms at room temperature, utilizing a multilayer FeSe film grown by MBE as the host material \cite{GiaoJPSJ2017}. {\it In-situ} ARPES measurements revealed that a pristine multilayer FeSe film is a low-carrier semimetal with small hole and electron Fermi surfaces at the Brillouin zone center and corner, respectively [Fig. 5(e)], similar to bulk FeSe \cite{MaletzPRB2014, NakayamaPRL2014}. On the other hand, after Li deposition, the electron Fermi surface expanded significantly [Fig. 5(f)], confirming a substantial electron doping into the FeSe layer. The Brillouin zone also expanded with Li deposition [compare gray vertical lines in Figs. 5(e) and 5(f)], as supported by a narrowing of the distance between the adjacent RHEED streaks. This suggests successful Li insertion which results in the relaxation of the tensile strain induced in pristine films by the lattice mismatch with the substrate \cite{TanNatMater2013, GiaoPRB2017}. Moreover, high-resolution measurements of the temperature-dependent superconducting gap revealed the occurrence of superconductivity with $T_{\rm c}$ of approximately 40 K, substantially higher than $T_{\rm c}$ of 9 K in pristine bulk FeSe. These findings demonstrate the ability of topotactic alkali-metal insertion to enhance $T_{\rm c}$ of FeSe.

\section{Removal of atoms}
\subsection{Graphene on SiC}
Besides the insertion of atoms, the removal of atoms via TCR also plays a key role to control electronic properties. This is also the case for epitaxial graphene grown on hydrogen-terminated SiC(0001) substrate \cite{RiedlPRL2009}. Epitaxial graphene grown on SiC can be used for device applications because relatively large sample can be obtained compared to the one obtained with the exfoliation method. However, it is known to be carrier doped from the buffer layer \cite{RiedlPRB2007}, limiting the device application. This buffer layer consists of carbon atoms with a graphene-like honeycomb structure covalently bonded with the SiC substrate \cite{MattauschPRL2007}, leading to the absence of $\pi$ band as demonstrated in Fig. 6(a). To realize quasi-free-standing graphene by suppressing carrier doping from the buffer layer, hydrogen atoms can be inserted between the buffer layer and the top of the SiC substrate, by annealing SiC at $\sim$600 $^\circ$C under hydrogen atmosphere \cite{RiedlPRL2009}. With this process, the buffer layer can be converted to monolayer graphene since the bonding between the buffer layer and the SiC substrate can be terminated, as is evident from the observation of the $\pi$ band in Fig. 6(b). Interestingly, the Dirac point is now located at slightly above $E_{\rm F}$ due to the tiny hole doping associated with the insertion of small amount of hydrogen molecules. After heating the sample up to 700 $^\circ$C, this hole doping effect vanishes and the ideal charge-neutral band structure with the Dirac point at $E_{\rm F}$ was realized [Fig. 6(c)]. When the sample was annealed with increased annealing temperature to $\sim$900 $^{\circ}$C, the hydrogen atoms were completely removed and the graphene layer was recoupled with the SiC substrate. This leads to the disappearance of $\pi$ bands [Fig. 6(d)] and the conversion of monolayer graphene to the buffer layer, signifying the reproducible nature of this TCR process \cite{RiedlPRL2009}. 

\begin{figure}
\begin{center}
\includegraphics[width=3.9in]{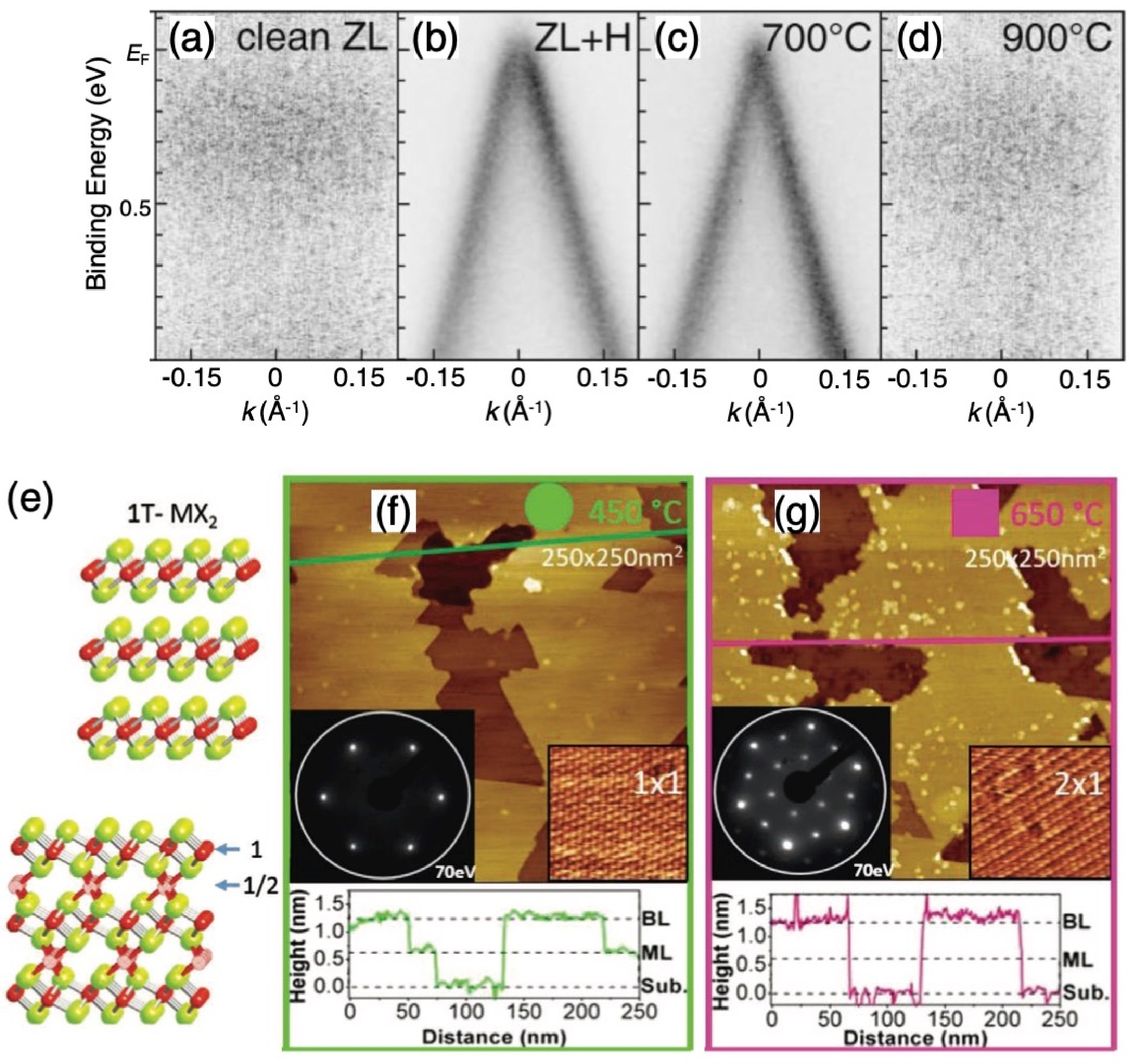}
\caption{(a)-(d) Experimental band dispersion around the K point of graphene Brillouin zone for (a) as-grown zero-layer (ZL) graphene (i.e. buffer layer) on SiC(0001), (b) after hydrogen treatment, and (c)-(d) subsequent annealing steps \cite{RiedlPRL2009}. (e) Crystal structure of 1T-MX$_2$ (top) and (bottom) atom-inserted 1T-MX$_2$. (f), (g) STM topography of as-grown (green panel) and post-annealed (magenta panel) VSe$_2$ thin film. Insets show LEED pattens and magnified STM images. Bottom panels show line profiles obtained along green and purple solid lines \cite{BonillaAMI2020}.}
 \end{center}
\end{figure}

\subsection{Transition-metal chalcogenides}
  As mentioned in section 2.3, the insertion of atoms in TMDCs is known to lead to different crystal phases with different chemical compositions. When transition-metal atoms are inserted into the van der Waal gap, they form covalent bonds with chalcogen atoms of the adjacent MX$_2$ layers, leading to a variety of crystal structures and characteristic physical properties depending on the amount and environment of the inserted transition-metal atoms. Such special type of TMDCs is often obtained by controlling the growth conditions \cite{FujisawaAdvMater2023} or by transition-metal insertion (section 2.3). Alternatively, it can be also obtained by the removal of chalcogen atoms via post-annealing \cite{BonillaAMI2020, LasekAPR2022}. Specifically, high-temperature post-annealing of MX$_2$ thin films induces a partial removement of chalcogen atoms (due to the higher vapor pressure of chalcogens compared to transition metals) and rearrangements of atoms, resulting in the self-insertion of excess transition-metal atoms into the van der Waals gap [Fig. 6(e)]. This TCR process facilitates the fabrication of TMDC compounds with a wide range of transition-metal and chalcogen atom ratios. Figures 6(f) and 6(g) display the STM topography of an MBE-grown VSe$_2$ thin film obtained at different post-growth annealing temperatures. The surface of the film annealed at 450 $^{\circ}$C exhibits a (1 $\times$ 1) periodicity associated with the growth of 1T-VSe$_2$ phase [top panel of Fig. 6(f)]. In contrast, the film annealed at 650 $^{\circ}$C displays a (2 $\times$ 1) periodicity in wide surface region, and this is attributed to the formation of V$_3$Se$_4$ phase [Fig. 6(g)]. Interestingly, the V$_3$Se$_4$ compound fabricated through the post-annealing exhibits electronic phases distinct from those grown directly by the MBE technique \cite{NakanoNanoLett2019}. The post-annealing was also applied to other TMDCs \cite{LasekAPR2022} such as Cr$_{1+\delta}$Te$_2$ thin film that exhibits ferromagnetism. The annealing increases the amount of self-intercalated Cr atoms in Cr$_{1+\delta}$Te$_2$, resulting in the modification of electronic states and magnetic properties \cite{LasekAPR2022}. Therefore, the annealing and resultant effective removal of chalcogen atoms serves as a practical approach to investigate the structural phases and physical properties in the TMDCs.

\subsection{Vacancy ordered FeSe}

\begin{figure}
\begin{center}
\includegraphics[width=4.5in]{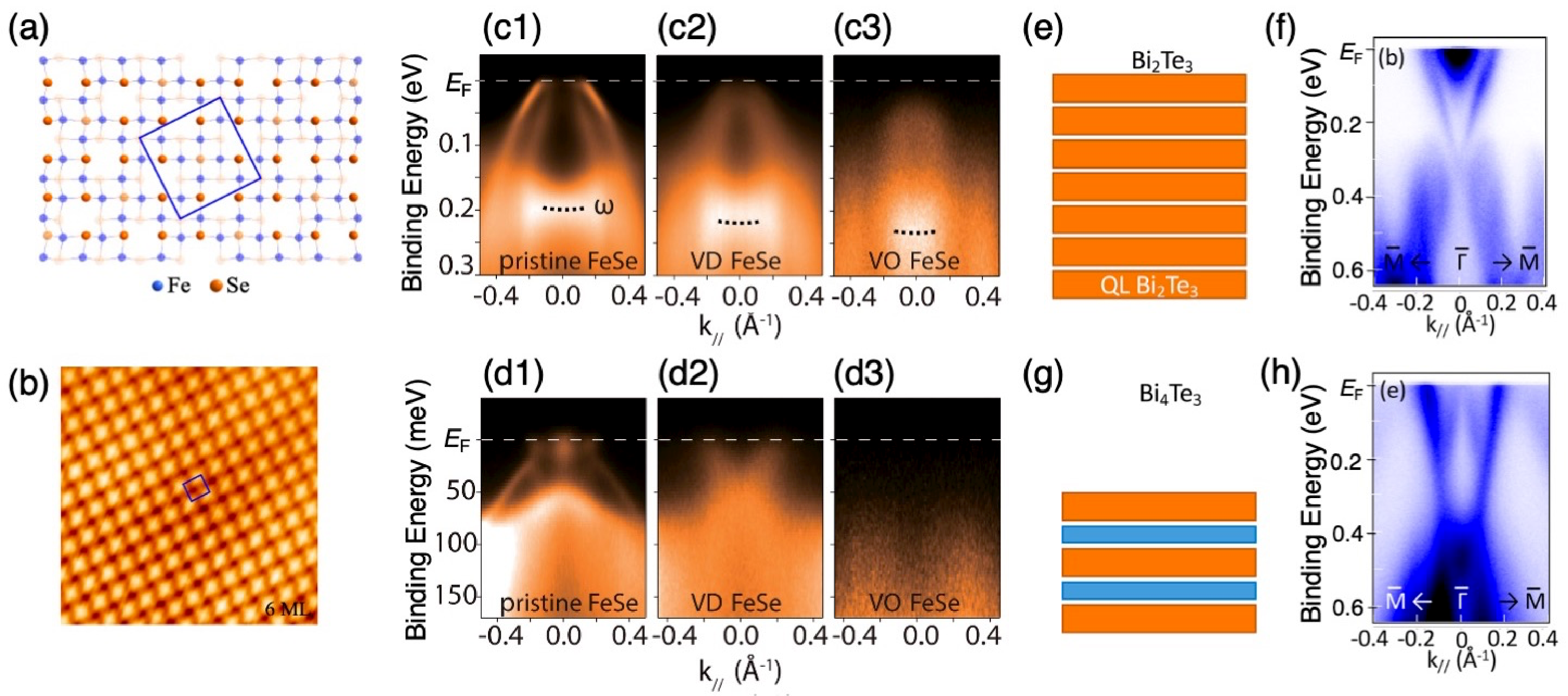}
\caption{(a) Sketched lattice model of Fe-vacancy Fe$_4$Se$_5$ phase \cite{YuPRM2020}. (b) High-resolution STM topography of $\sqrt{5} \times \sqrt{5}$ superstructure in the vacancy ordered FeSe films. (c)-(d) ARPES intensity plots at the $\Gamma$ and M points, respectively, for as-grown, vacancy disordered, and vacancy ordered FeSe thin films \cite{FangPRB2016}. (e), (f) Schematic crystal sheets of Bi$_2$Te$_3$ and corresponding band dispersion of pristine Bi$_2$Te$_3$ along the $\Gamma$M cut \cite{KusakaAPL2022}. (g), (h), Same as (e) and (f) but for the Bi$_4$Te$_3$ superlattice.}
 \end{center}
\end{figure}

The removal of atoms also plays a crucial role to the superconductivity in FeSCs. As described in section 2.4, bulk FeSe exhibits superconductivity with $T_{\rm c}$ of 9 K in an undoped sample \cite{HsuPNAS2008} and a higher $T_{\rm c}$ of around 40 K after electron doping \cite{GuoPRB2010, WangPRB2011, FangEPL2011, YingSciRep2012, ScheidtEPJB2012, KrztonJPCM2012, BurrardNatMater2013, NojiPhysC2014, LuNatMater2014}. While earlier studies suggested an antiferromagnetic metallic state in FeTe as the parent phase for the superconductivity \cite{YehEPL2008, FangPRB2008, MizuguchiJPSJ2009}, recent study pointed out that the parent phase may be an insulator with the chemical composition of Fe$_4$Se$_5$ characterized by a $\sqrt{5} \times \sqrt{5}$ Fe vacancy order \cite{ChenPNAS2014} [Fig. 7(a)]. The vacancy-ordered phase with an insulating ground state was also found in alkali-metal-inserted FeSe (i.e., K$_2$Fe$_4$Se$_5$) \cite{WangPRB2011, BaoCPL2011, YePRL2011, WeiNatPhys2012}, which is closely related to the superconducting counterparts (K$_{2+x}$Fe$_{4+y}$Se$_5$). In these systems, superconductivity emerges as the Fe vacancy becomes disordered. Therefore, there is a huge demand to elucidate the evolution of the electronic states from vacancy-ordered FeSe (i.e., Fe$_4$Se$_5$) to vacancy-disordered states. In this context, the TCR has proven to be powerful. It was discovered that annealing as-grown FeSe films in UHV at low temperatures (ranging from 550 $^{\circ}$C to 250 $^{\circ}$C) for a few hours resulted in the formation of vacancy-disordered FeSe, and additional annealing at 250 $^{\circ}$C for 16 h successfully generated vacancy-ordered FeSe with the characteristic $\sqrt{5} \times \sqrt{5}$ superstructure [Fig. 7(b)] \cite{YuPRM2020}. Furthermore, it was demonstrated that the vacancy-ordered FeSe film can be reversely converted to vacancy-disordered FeSe by gentle annealing. Fang {\it et al.} conducted {\it in-situ} ARPES measurements \cite{FangPRB2016} and revealed that, in contrast to a pristine FeSe film being a compensated semimetal \cite{TanNatMater2013, MiyataNatMater2015}, the vacancy-disordered FeSe film is an electron-doped metal, and the vacancy-ordered FeSe film is an insulator [Figs. 7(c) and 7(d)]. Recent spin-resolved STM measurements \cite{ZhangAdvMater2023} further suggested antiferromagnetic ordering in the vacancy-ordered FeSe film. Considering that the vacancy-disordered FeSe film is likely a superconductor, the electronic phase diagram in FeSe film is similar to that of high-$T_{\rm c}$ cuprate superconductors, in which superconductivity is realized through a transition from an antiferromagnetic insulator to a metal by carrier doping. The TCR allows systematic studies of this intriguing electronic phase diagram and the electronic states therein to advance our understanding of the superconducting mechanism in FeSe.

\subsection{Topological superlattice}
The removal of atomic layers in TI is known to be helpful to strongly modulate topological properties. The basic structural unit of prototypical TI, Bi$_2$Te$_3$, is a QL composed of the Te-Bi-Te-Bi-Te atomic layers [orange rectangle in Fig. 7(e)], and QL layers are periodically stacked along c-axis. It was reported that annealing Bi$_2$Te$_3$ leads to the removement of Te layers in the crystal \cite{KusakaAPL2022}, and this process is regarded as a TCR. As a result of the Te removement, two distinct stoichiometric phases were obtained. One is the Bi$_1$Te$_1$ phase having a basic unit of 1QL Bi$_2$Te$_3$ - 1 bilayer (BL) Bi - 1QL Bi$_2$Te$_3$ and the other is the Bi$_4$Te$_3$ phase with the unit of 1QL Bi$_2$Te$_3$ - 1BL Bi [Fig. 7(g)]. These structures can be regarded as a topological superlattice consisting of an alternate stacking of 1QL and 1BL units, and such superlattices can be grown also in the bulk form \cite{BosPRB2007}. As shown in Fig. 7(h), the ARPES measurements of Bi$_4$Te$_3$ clarified Dirac-cone-like feature which is heavily electron doped compared to pristine Bi$_2$Te$_3$ [Fig. 7(f)], together with a holelike band crossing $E_{\rm F}$ that is likely associated with 1BL Bi \cite{NabokPRM2022}. Topological nature of these superlattices was discussed in terms of the dual topological state characterized by a weak/strong TI phase with Z$_2$ topological invariant and a topological crystalline insulator phase with mirror Chern number \cite{KusakaAPL2022}.

\section{Replacement of atoms}
\subsection{Superconducting bilayer graphene}

\begin{figure}
\begin{center}
\includegraphics[width=3.9in]{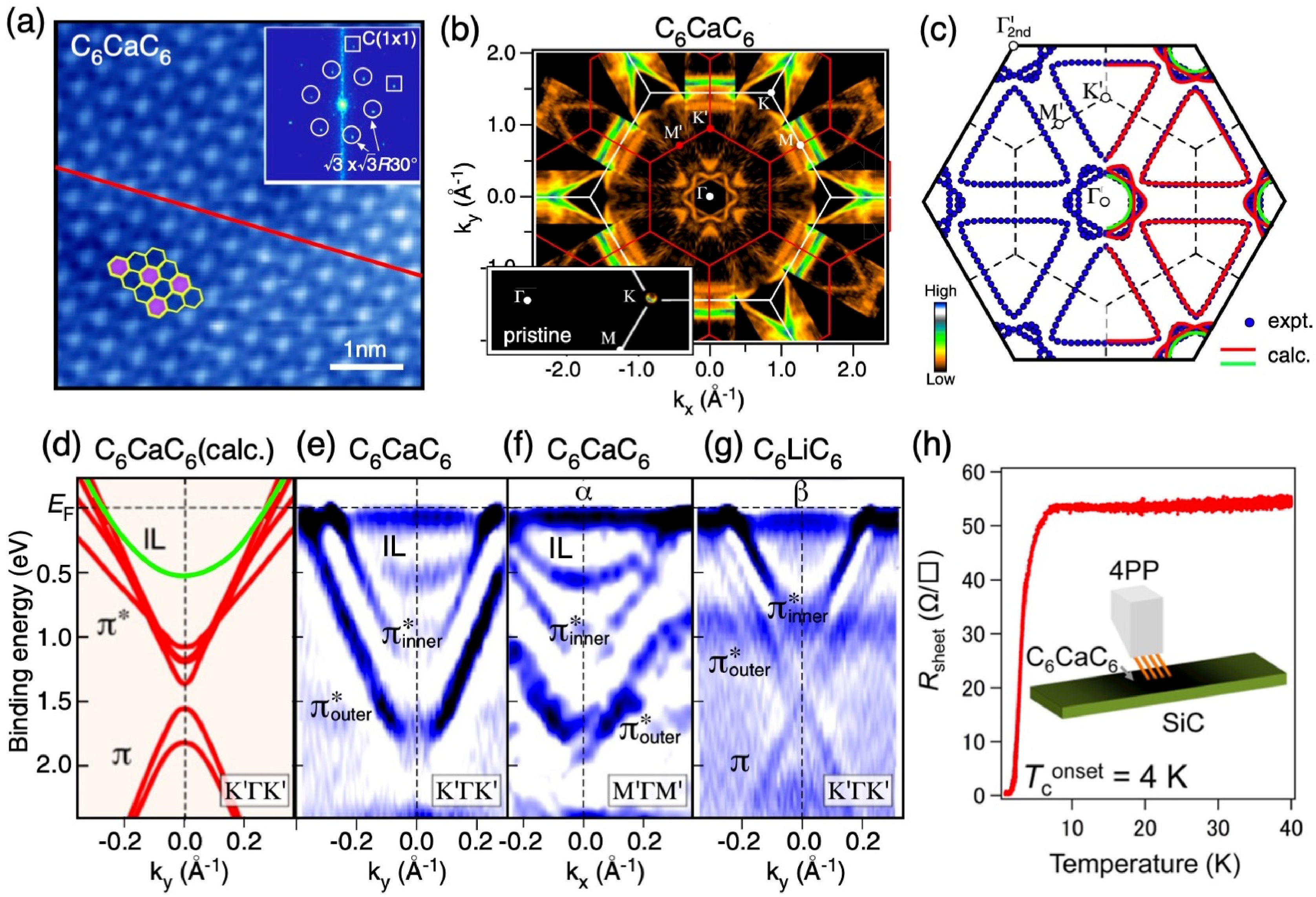}
\caption{(a) STM image of Ca-inserted bilayer graphene, C$_6$CaC$_6$, fabricated by TCR. (b) Fermi-surface mapping of C$_6$CaC$_6$ and bilayer graphene/6H-SiC (inset). White and red lines correspond to the Brillouin zones of graphene and C$_6$CaC$_6$, respectively. (c) Comparison of Fermi wave vectors obtained by the experiment and calculation. (d) Calculated band structure of C$_6$CaC$_6$ along the $k_y$ cut. (e), (f) Second-derivative of the ARPES intensity along the $k_y$ and $k_x$ cuts, respectively, for C$_6$CaC$_6$ \cite{KanetaniPNAS2012}. (g) Same as (e) but for C$_6$LiC$_6$. (h) Temperature dependence of sheet resistance for C$_6$CaC$_6$ obtained by 4-point-probe measurement \cite{IchinokuraACSNano2016}.}
 \end{center}
\end{figure}

Besides the insertion and removal of atoms, the replacement of atoms is widely applied to realize exotic physical properties in atomic-layer materials. This is highlighted by atom-inserted bilayer graphene. Kanetani {\it et al.} demonstrated that Li atoms in C$_6$LiC$_6$ (see section 2.1) can be completely replaced with Ca atoms by annealing C$_6$LiC$_6$ at 450 $^{\circ}$C after Ca deposition. The formation of Ca-inserted bilayer graphene, C$_6$CaC$_6$ \cite{KanetaniPNAS2012}, is suggested from the comparison of the STM images between C$_6$LiC$_6$ and C$_6$CaC$_6$ shown in Figs. 4(c) and 8(a). Although the local atomic position of the bright spot in the STM images looks different between C$_6$LiC$_6$ and C$_6$CaC$_6$, their Fourier-transform images commonly show similar spots with the periodicity of $\sqrt{3} \times \sqrt{3} R30^{\circ}$. One can recognize in Fig. 8(b) that the Fermi surface topology of C$_6$CaC$_6$ is apparently different from that of the pristine sample (inset), and shows a good agreement with the calculated Fermi surface for free-standing C$_6$CaC$_6$ \cite{MazinPML2010} [Fig. 8(c)]. This establishes that Ca atoms are indeed inserted into graphene layers in well-ordered manner. As described in section 2.1, the near-$E_{\rm F}$ band structure around the $\Gamma$ point consists of folded $\pi$/$\pi^*$ bands in C$_6$LiC$_6$ (note that there are inner and outer $\pi$/$\pi^*$ bands) [Fig. 8(g)]. On the other hand, besides the $\pi$/$\pi^*$ bands, a parabolic band with a bottom at 0.5 eV can be recognized in C$_6$CaC$_6$ [Figs. 8(e) and 8(f)]. Experimental band dispersion of this parabolic band shows a good agreement with the DFT-derived band dispersion associated with the interlayer band showing nearly-free-electron-like character [green curve in Fig. 8(d)] \cite{MazinPML2010, JishiASTP2011, PosternakPRL1983}. Since the interlayer band in the bulk GICs was discussed to be directly responsible for the occurrence of superconductivity \cite{PosternakPRL1983, OhnoJPSJ1979}, it is expected that the interlayer electrons clarified by ARPES also induce superconductivity in C$_6$CaC$_6$. To examine this point, Ichinokura {\it et al.} carried out temperature-dependent electrical resistivity measurement with the {\it in-situ} micro four-point probe method for both C$_6$LiC$_6$ and C$_6$CaC$_6$. As shown in Fig. 8(h), they observed zero electrical resistance at around $T_{\rm c} \sim$ 2 K associated with superconductivity in C$_6$CaC$_6$, in contrast to the absence of superconductivity in C$_6$LiC$_6$ \cite{IchinokuraACSNano2016}. This result indicates that atomic replacement utilizing the TCR is a key to realize superconducting bilayer graphene. It is noted that the $T_{\rm c}$ of Ca-inserted bilayer graphene can be further enhanced up to $\sim$6 K upon additional insertion of Ca atoms between graphene and SiC due to the extra electron doping \cite{ToyamaACSNano2022}.

\subsection{Atomic-layer FeSC}
In addition to the 40-K superconductivity triggered by electron doping and the intriguing phase diagram reminiscent of high-$T_{\rm c}$ cuprate superconductors, FeSe is also attracting significant interest due to the discovery of superconductivity in a monolayer film grown on SrTiO$_3$ (monolayer FeSe/SrTiO$_3$), with a $T_{\rm c}$ value as high as 65 K \cite{TanNatMater2013, MiyataNatMater2015, WangCPL2012, HeNatMater2013, ShiNatCommun2016}, which stands as the highest among FeSCs. A widely debated scenario to explain such high $T_{\rm c}$ in monolayer FeSe/SrTiO$_3$ is a strong coupling between SrTiO$_3$ phonons and FeSe electrons at the interface, which is evident from the appearance of phonon shake-off subbands in ARPES measurements \cite{LeeNature2014, RebecPRL2017, SongNatCommun2019}. To further test this scenario, Shigekawa {\it et al.} performed a comparative study on a monolayer film of FeS \cite{ShigekawaPNAS2019}, which is a superconductor with $T_{\rm c}$ (4.5 K) comparable to that of isostructural FeSe in its bulk crystal form \cite{LaiJACS2015}. Obtaining high-quality tetragonal FeS thin films using MBE has been challenging, primarily due to (i) the high vapor pressure of S and (ii) the metastable nature of the tetragonal FeS compared to the hexagonal phase. Nevertheless, they have overcome the difficulties by utilizing tetragonal monolayer FeTe/SrTiO$_3$ as the starting material and heated it at 270 $^{\circ}$C for 5 min while irradiating S molecular beam to replace Te with S via TCR, followed by post-annealing at 530 $^{\circ}$C for 1 h. The successful sulfurization and the tetragonal structure of the obtained monolayer FeS/SrTiO$_3$ were confirmed by STEM-EDX, LEED, XPS, and ARPES measurements. In terms of the electronic structure, monolayer FeS/SrTiO$_3$ displayed sharp quasiparticle bands, indicative of its high-quality nature [Fig. 9(b)]. Unlike the presence of a holelike Fermi surface at the $\Gamma$ point in monolayer FeTe/SrTiO$_3$ [Fig. 9(a)], monolayer FeS/SrTiO$_3$ exhibited only electronlike Fermi surfaces at the M point [denoted as $\gamma / \delta$ in Fig. 9(b)] due to electron charge transfer from SrTiO$_3$, similar to monolayer FeSe/SrTiO$_3$. Additionally, ARPES intensity displayed a subband ($\gamma$') near the M point in monolayer FeS/SrTiO$_3$, which was well reproduced by a $\sim$100-meV downward shift of the $\gamma$ band - an energy shift close to the energy of optical phonon modes in SrTiO$_3$ [Figs. 9(c)-9(e)]. These characteristics, similar to those of the replica band produced by the phonon shake-off effect in monolayer FeSe/SrTiO$_3$ \cite{LeeNature2014, RebecPRL2017, SongNatCommun2019}, indicated a strong interfacial electron-phonon coupling in monolayer FeS/SrTiO$_3$. The interfacial origin of the $\gamma$' band was supported by its absence in multilayer counterparts [Fig. 9(f)]. Surprisingly, despite observing the interfacial electron-phonon coupling (and the electron doping), Shigekawa {\it et al.} reported the absence of high-$T_{\rm c}$ superconductivity in monolayer FeS/SrTiO$_3$. This intriguing finding suggests that the cross-interface electron-phonon coupling enhances $T_{\rm c}$ only when it cooperates with the pairing interaction inherent to the superconducting layer, offering a key insight to explore heterointerface high-$T_{\rm c}$ superconductors.

\begin{figure}
\begin{center}
\includegraphics[width=3.9in]{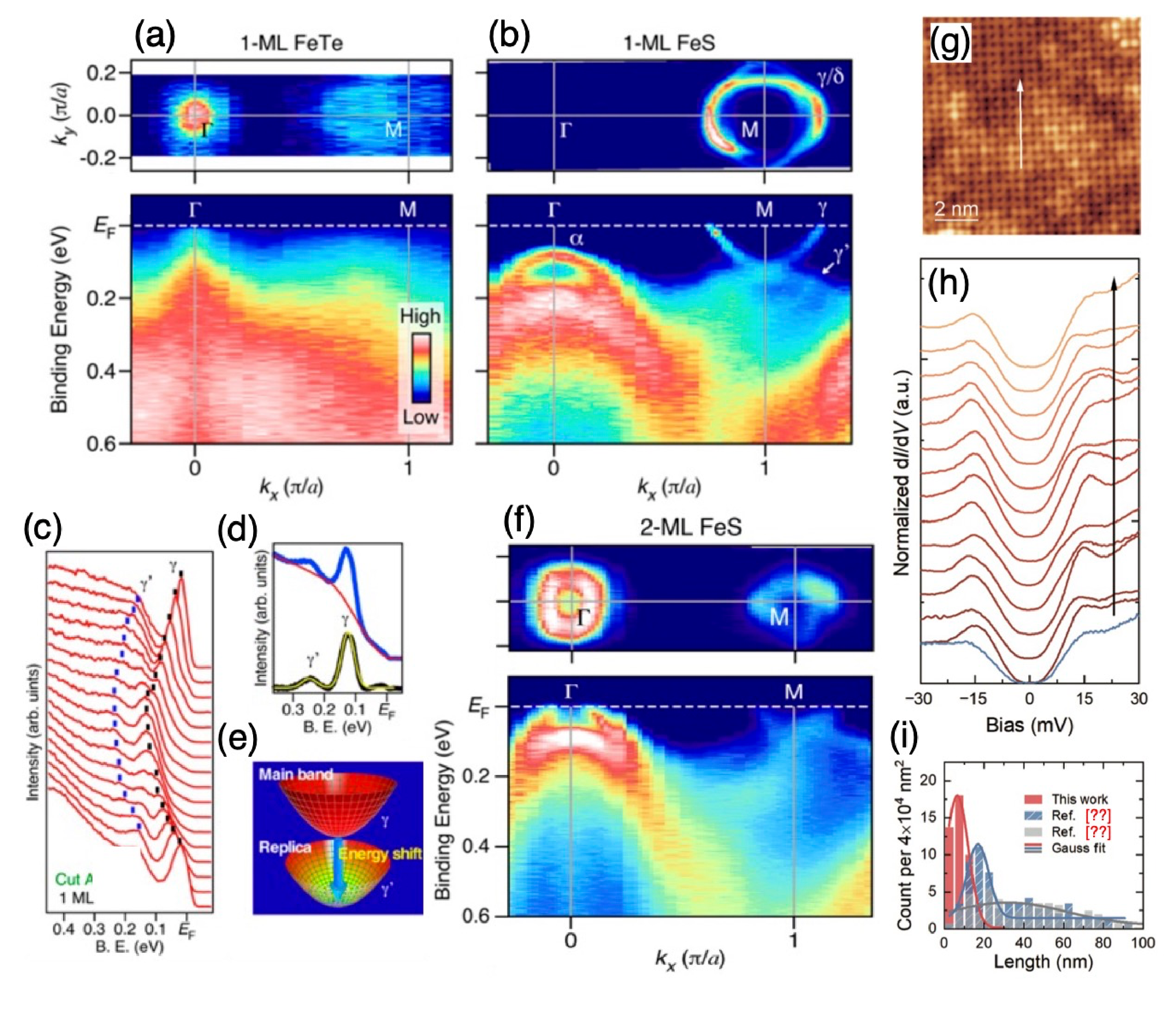}
\caption{(a), (b) ARPES intensity plots for monolayer FeTe and monolayer FeS films as a function of 2D wave vectors, respectively. (c) Near-$E_{\rm F}$ ARPES spectra for monolayer FeS thin film along the $\Gamma$M$\Gamma$ cut. (d) Numerical fitting for the ARPES spectrum. (e) Schematics for replica bands. (f) ARPES intensity plots for 2 monolayer FeS thin film as a function of 2D wave vectors \cite{ShigekawaPNAS2019}. (g) STM topography of FeSe film converted from the FeTe film. (h) A set of tunneling spectra measured along white arrow shown in (g). (i) Statistics of number and size of line defects \cite{WeiNanoRes2023, ZhangPRB2014, GeNanoLett2019}. 
}
 \end{center}
\end{figure}

In addition to the sulfurization of FeTe, Shigekawa {\it et al.} demonstrated successful conversion from FeTe to FeSe and from FeSe to FeS, emphasizing the importance of large electronegativity differences for successful topotactic replacement of chalcogen atoms \cite{ShigekawaPNAS2019}. Furthermore, Wei {\it et al.} recently reported partial topotactic replacement of Te with Se \cite{WeiNanoRes2023}. They conducted STM measurements and observed sharper coherence peaks in the superconducting state of the obtained Fe(Se,Te)/SrTiO$_3$ [Fig. 9(h)] compared to the film prepared by conventional co-evaporation of Fe, Se, and Te atoms. They also revealed fewer defects in the film obtained by TCR [Figs. 9(g) and 9(i)], indicating a high sample quality. These studies highlight the advantage of TCR for synthesizing metastable chalcogenides, with promising applications in various 2D chalcogenides such as the TMDCs and topological materials.

\subsection{Charge-density-wave TMDC compounds}

\begin{figure}
\begin{center}
\includegraphics[width=3.9in]{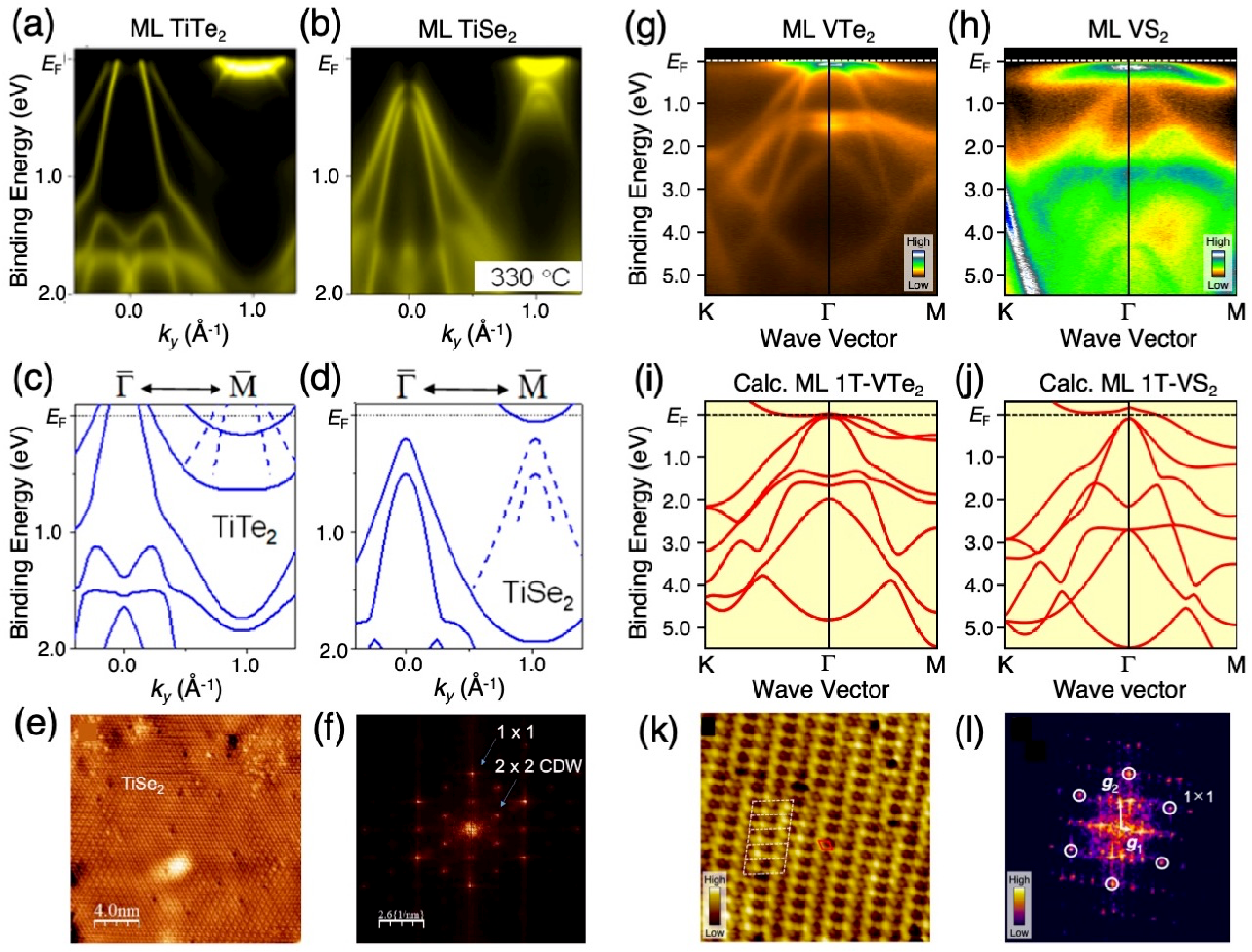}
\caption{(a), (b) ARPES intensity plots for as grown monolayer TiTe$_2$ and monolayer TiSe$_2$ obtained by selenization. (c), (d) Calculated band structure of monolayer 1T-TiTe$_2$ and monolayer 1T-TiSe$_2$ obtained by the DFT calculations. (e) A topographic image taken from TiSe$_2$ region in TiTe$_{1.5}$Se$_{0.5}$. (f) Fourier transform image of (e). The (1 $\times$ 1) spots are derived from the normal phase, while the (2 $\times$ 2) spots are from the CDW phase \cite{ChenACSNano2019}. (g), (h) ARPES-intensity plots of monolayer VTe$_2$ and VS$_2$ films, respectively, measured along the K$\Gamma$M cut. monolayer VS$_2$ is converted from monolayer VTe$_2$ via TCR. (i), (j) Calculated band structure of monolayer 1T-VTe$_2$ and monolayer 1T-VS$_2$ obtained with the DFT calculations. (k) High-resolution STM image for monolayer VS$_2$ thin film in the region of 10 $\times$ 10 nm$^2$. White broken rhombus indicates the unit cell of $\sqrt{21} R 10.9^{\circ} \times \sqrt{3} R30^{\circ}$ superstructure, whereas the red solid rhombus corresponds to the original (1 $\times$ 1) unit cell. (l) Fourier transform image of (k). White arrows indicate primitive reciprocal lattice vectors {\bf g}$_1$ and {\bf g}$_2$ for the CDW periodicity \cite{Kawakaminpj2DMA2023}.
}
 \end{center}
\end{figure}

The CDW has long been attracted tremendous attention in layered materials not only due to the fundamental interest on its driving mechanism but also due to its intriguing competition or coexistence with the superconductivity and magnetism, as exemplified by high-$T_{\rm c}$ cuprate superconductors. Among layered materials, atomic-layer TMDCs are a promising platform to investigate CDW, partly owing to the enhancement of electron-phonon coupling in the 2D limit. A variety of CDW phases with different periodicities were reported in monolayer TMDCs, and such CDW was found to be further controlled by the strain from the substrate and carrier doping. Atomic replacement utilizing the TCR also serves as a valuable method for controlling the CDW phase \cite{Kawakaminpj2DMA2023, ChenACSNano2019}. Chen {\it et al.} have successfully obtained monolayer TiSe$_2$ from monolayer TiTe$_2$ by selenization \cite{ChenACSNano2019} as can be seen from the ARPES intensity plots along the $\Gamma$M cut measured before and after the replacement [Figs. 10(a) and 10(b), respectively]. The former exhibits a semi-metallic band dispersion characterized by hole- and electron-bands crossing $E_{\rm F}$ centered at the $\Gamma$ and M points, respectively. On the other hand, the latter exhibits the insulating nature of the hole band at the $\Gamma$ point while keeping the metallic states with the electron bands crossing $E_{\rm F}$ around the M point. Remarkably, it also shows a replica hole band around the M point associated with the formation of 2 $\times$ 2 CDW \cite{ChenNatCommun2015, SugawaraACSNano2016}. These features show excellent agreement with those of monolayer TiSe$_2$ fabricated by the MBE method without using TCR, and also with the DFT-derived band structure for free-standing monolayer TiSe$_2$ assuming 2 $\times$ 2 band folding [Figs. 10(c) and 10(d)]. The formation of 2 $\times$ 2 CDW is also corroborated by the STM experiments [Figs. 10(e) and 10(f)].

While transition-metal diselenides can be fabricated by the MBE method, it is rather difficult to obtain transition-metal disulfides by the MBE method alone. The TCR works well in this respect, as exemplified by the fabrication of monolayer VS$_2$ thin film \cite{Kawakaminpj2DMA2023}. Figures 10(g) and 10(h) display ARPES intensity for as-grown monolayer VTe$_2$ and monolayer VS$_2$ thin film obtained with the sulfurization of VTe$_2$ via TCR, respectively. The former exhibits relatively narrow metallic bands around $E_{\rm F}$ and dispersive hole bands centered at the $\Gamma$ point that are attributed to the V 3$d$ and Te 5$p$ orbitals, respectively [Fig. 10(g)]. In contrast, sulfurization leads to a significant change in the valence bands [Fig. 10(h)]. In particular, the hole band moves downward after sulfurization; this is associated with the lower energy level of the S 3$p$ atomic orbital compared to the Te 5$p$ orbital, as expected from the DFT calculations [Figs. 10(i) and 10(j)]. Interestingly, monolayer VS$_2$ exhibits a superstructure associated with a $\sqrt{21}R10.9 ^{\circ} \times \sqrt{3}R30 ^{\circ}$ CDW, as evident from the STM image and corresponding Fourier transform image in Figs. 10(k) and 10(l), respectively. Remarkably, the observed superstructure is distinct from that of monolayer VTe$_2$ \cite{WangPRB2019}. This result certainly suggests that sulfurization via TCR serves as an excellent means to control the CDW phase of atomic-layer TMDCs.

\section{Summary and Outlook}
In summary, the TCR method that involves the insertion, removal, or replacement of atoms was shown to be a key to realize and control exotic electronic properties. The advantages of TCR are particularly pronounced for the atomic-layer materials owing to their high surface-to-volume ratio. Therefore, the combination of the MBE and TCR methods as well as the visualization of band structure via advanced surface-sensitive spectroscopy techniques is highly valuable for exploring functional atomic-layer materials. It is emphasized that spectroscopic investigations combined with the TCR are still in the early stages, and the application of this technique to a variety of systems is highly encouraged. For example, perovskite materials fabricated by the TCR are a promising research target due to their unique properties such as magnetism and superconductivity \cite{KageyamaNatCommun2018}, although the investigation of their electronic states through {\it in-situ} spectroscopies has yet to be conducted. In addition, other low-dimensional systems such as MXeane and quasi-1D materials may be plausible candidates to explore new functionalities through TCR and spectroscopic investigations. It is also important to utilize other reactants to further expand accessible material families. For example, halogen elements with high electronegativity can be used for synthesizing halogen compounds like BiTeBr, which shows giant Rashba spin splitting \cite{HajraACSNano2020, VaneyNatCommun2022}. Nitrogen and oxygen atoms \cite{Xia2DMater2018, MairoserNatCommun2015, KhareAdvMater2017, YangCIE2020, WangAdvMat2022} could also broaden the possibilities for material exploration. In spectroscopy viewpoint, ARPES with micro- or nano-focused beam spot would enrich the potential for material investigation, even for the samples in which large single crystals can hardly be synthesized. Real-time monitoring of band structures and core levels during the TCR is also valuable for understanding the actual reaction process.

\section{Data availability statement}
All data that support the findings of this study are included within the article.

\section{Acknowledgments}
This work was supported by JST-CREST (no. JPMJCR18T1), JST-PRESTO (no. JPMJPR20A8), Grant-in-Aid for Scientific Research (JSPS KAKENHI Grant Numbers JP18H01821, JP20H01847, JP21H01757, JP21H04435, and JP22J13724), KEK-PF (Proposal No. 2020G669, 2021S2-001, and 2022G007), Foundation for Promotion of Material Science and Technology of Japan, and World Premier International Research Center, Advanced Institute for Materials Research. T. K. acknowledges support from GP-Spin at Tohoku University and JSPS.

\section{ORCID iD}
Tappei Kawakami; https://orcid.org/0000-0003-4643-2805\\
Kosuke Nakayama; https://orcid.org/0000-0003-2462-2253\\
Katsuaki Sugawara; https://orcid.org/0000-0003-2926-9436\\
Takafumi Sato; https://orcid.org/0000-0002-4544-5463\\

\newpage


\begin{thebibliography}{50}
\bibitem{NovoselovNature2005} Novoselov K {\it et al} 2005 {\it Nature} {\bf 438} 197
\bibitem{KonigScience2007} K\"{o}nig M {\it et al} 2007 {\it Science} {\bf 318} 766
\bibitem{XiaNatPhys2009} Xia Y {\it et al} 2009 {\it Nat. Phys.} {\bf 5} 398
\bibitem{CaoNature2018} Cao Y {\it et al} 2018 {\it Nature} {\bf 556} 43
\bibitem{DengScience2020} Deng Y {\it et al} 2020 {\it Science} {\bf 367} 895
\bibitem{TranquadaNature1995} Tranquada J M {\it et al} 1995 {\it Nature} {\bf 375} 561
\bibitem{KamiharaJACS2008} Kamihara Y {\it et al} 2008 {\it J. Am. Chem. Soc.} {\bf 130} 3296
\bibitem{ChuangScience2010} Chuang T M {\it et al} 2010 {\it Science} {\bf 327} 181
\bibitem{XiNatPhys2016} Xi X {\it et al} 2016 {\it Nat. Phys.} {\bf 12} 139
\bibitem{GrubiAdvSci2023} Grubi\v{s}i\'{c}-\v{C}abo A {\it et al} 2023 {\it Adv. Sci.} {\bf 10} 2301243
\bibitem{ZhangAdvSci2022} Zhang Z {\it et al} 2022 {\it Adv. Sci.} {\bf 9} 2105201
\bibitem{DeependraAdvMater2022} Liang Y {\it et al} 2019 {\it Adv. Mater.} {\bf 31} 1901964
\bibitem{KageyamaNatCommun2018} Kageyama H {\it et al} 2018 {\it Nat. Commun.} {\bf 9} 772
\bibitem{CahenEJIC2019} Cahen S {\it et al} 2019 {\it Eur. J. Inorg. Chem.}  {\bf 2019} 4798
\bibitem{SoodNatRevMat2021} Sood A {\it et al} 2021 {\it Nat. Rev. Mater.}  {\bf 6} 847
\bibitem{LohNanoRes2021} Loh L {\it et al} 2021 {\it Nano Res.} {\bf 14} 1668
\bibitem{LamACSNano2022} Lam D, Lebedev D, and Hersam M C 2022 {\it ACS Nano} {\bf 16} 7144
\bibitem{ZhangNatNanotech2023} Zhang K {\it et al} 2023 {\it Nat. Nanotech.} {\bf 18} 448
\bibitem{LotgeringJINC1959} Lotgering F K 1959 {\it J. Inorg. Nucl. Chem.} {\bf 9} 113
\bibitem{ShannonNature1964} Shannon R, Rossi R, 1964 {\it Nature} {\bf 202} 1000.
\bibitem{VolpeCRSE1985} Volpe L and Boudart M 1985 {\it Catal. Rev. Sci. Eng.} {\bf 27} 515
\bibitem{FiglarzSSI1990} Figlarz M {\it et al} 1990 {\it Solid State Ionics} {\bf 43} 143
\bibitem{DelmasIJIM1999} Delmas C {\it et al} 1999 {\it Int. U. Inorg. Mater.} {\bf 1} 11
\bibitem{LiNature2019} Li D {\it et al} 2019 {\it Nature} {\bf 572} 624
\bibitem{KanetaniPNAS2012} Kanetani K {\it et al} 2012 {\it Proc. Natl. Acad. Sci. USA} {\bf 109} 19610
\bibitem{Kitteltext} Kittel C 1953 {\it Introduction to Solid State Physics} (John Wiley $\&$ Sons, Inc.)
\bibitem{NakataNatCommun2021} Nakata Y {\it et al} 2021 {\it Nat. Commun.} {\bf 12} 5873
\bibitem{ZhangNRMP2022} Zhang H {\it et al} 2022 {\it Nat. Rev. Methods Primers} {\bf 2} 54
\bibitem{ShigekawaPNAS2019} Shigekawa K {\it et al} 2019 {\it Proc. Natl. Acad. Sci. USA} {\bf 116} 24470
\bibitem{HiraharaNatCommun2020} Hirahara T {\it et al} 2020 {\it Nat. Commun.} {\bf 11} 4821
\bibitem{BaoPRL2021} Bao C {\it et al} 2021 {\it Phys. Rev. Lett.} {\bf 126} 206804
\bibitem{KusunoseAPL2022} Kusaka S {\it et al} 2022 {\it Appl. Phys. Lett.} {\bf 120} 173102
\bibitem{FujisawaAdvMater2023} Fujisawa H {\it et al} 2023 {\it Adv. Mater.} {\bf 35} 2207121
\bibitem{Kawakaminpj2DMA2023} Kawakami T {\it et al} 2023 {\it npj 2D Mater. Appl.} {\bf 7} 35
\bibitem{SugawaraAIPAdv2011} Sugawara K {\it et al} 2011 {\it AIP Adv.} {\bf 1} 022103
\bibitem{YagyuAPL2014} Yagyu K {\it et al} 2014 {\it Appl. Phys. Lett.} {\bf 104} 053115
\bibitem{AndersonRPM2017} Anderson N A {\it et al} 2017 {\it Phys. Rev. Mater.} {\bf 1} 054005
\bibitem{YajiPRL2019} Yaji K {\it et al} 2019 {\it Phys. Rev. Lett.} {\bf 122} 126403
\bibitem{BriggsNatMater2020} Briggs N {\it et al} 2020 {\it Nat. Mater.} {\bf 19} 637
\bibitem{RiedlPRL2009} Riedl C {\it et al} 2009 {\it Phys. Rev. Lett.} {\bf 103} 246804
\bibitem{BronoldAPA1991} Bronold M {\it et al} 1991 {\it Appl. Phys. A} {\bf 52} 171
\bibitem{BrauerRPB1997} Brauer H E {\it et al} 1997 {\it Phys. Rev. B} {\bf 55} 15
\bibitem{NakataPRM2019} Nakata Y {\it et al} 2019 {\it Phys. Rev. Mater.} {\bf 3} 071001 
\bibitem{BonillaAMI2020} Bonilla M {\it et al} 2020 {\it Adv. Mater. Interfaces} {\bf 7} 2000497
\bibitem{LasekAPR2022} Lasek K {\it et al} 2022 {\it Appl. Phys. Rev.} {\bf 9} 011409
\bibitem{Xia2DMater2018} Xia Y {\it et al} 2018 {\it 2D Mater.} {\bf 5} 041005
\bibitem{BarjaNatCommun2019} Barja S {\it et al} 2019 {\it Nat. Commun.} {\bf 10} 3382
\bibitem{ChenACSNano2019} Chen P {\it et al} 2019 {\it ACS Nano} {\bf 13} 5611
\bibitem{HorPRL2010} Hor Y S {\it et al} 2010 {\it Phys. Rev. Lett.} {\bf 104} 057001
\bibitem{HiraharaNanoLett2017} Hirahara T {\it et al} 2017 {\it Nano Lett.} {\bf 17} 3493
\bibitem{KusakaAPL2022} Kusaka S {\it et al} 2022 {\it Appl. Phys. Lett.} {\bf 120} 17
\bibitem{PhanJPS2017} Phan G N {\it et al} 2017 {\it J. Phys. Soc. Jpn.} {\bf 86} 033706
\bibitem{FangPRB2016} Fang Y {\it et al} 2016 {\it Phys. Rev. B} {\bf 93} 184503
\bibitem{WeiNanoRes2023} Wei Z {\it et al} 2023 {\it Nano Res.} {\bf 16} 1712
\bibitem{ZhangPRB2014} Zhang W H {\it et al} 2014 {\it Phs. Rev. B} {\bf 89} 060506(R)
\bibitem{GeNanoLett2019} Ge Z Z {\it et al} 2019 {\it Nano Lett.} {\bf 19} 2497
\bibitem{EbertARM1976} Ebert L B 1976 {\it Annu. Rev. Mater. Sci.} {\bf 6} 181
\bibitem{DresselhausAdvPhys1981} Dresselhaus M S and Dresselhaus G 1981 {\it Adv. Phys.} {\bf 30} 139
\bibitem{VirojanadaraPRB2010} Virojanadara C {\it et al} 2010 {\it Phys. Rev. B} {\bf 82} 205402 
\bibitem{LudbrookPNAS2015} Ludbrook B M {\it et al} 2015 {\it Proc. Natl. Acad. Sci. USA} {\bf 112} 11795 
\bibitem{HuempfnerAMI2022} Huempfner T {\it et al} 2022 {\it Adv. Mater. Interfaces}  {\bf 9} 2200585
\bibitem{ToyamaACSNano2022} Toyama H {\it et al} 2022 {\it ACS Nano} {\bf 16} 3582
\bibitem{McChesneyPRL2010} McChesney J L {\it et al} 2010 {\it Phys. Rev. Lett.} {\bf 104} 136803
\bibitem{WarmuthPRB2016} Warmuth J {\it et al} 2016 {\it Phys. Rev. B} {\bf 93} 165437
\bibitem{KShenJPCC2018} Shen K {\it et al} 2018 {\it J. Phys. Chem. C} {\bf 122} 21484
\bibitem{NandkishoreNatPhys2012} Nandkishore R {\it et al} 2012 {\it Nat. Phys.} {\bf 8} 158
\bibitem{LiNatMater2023} Li C {\it et al} 2023 {\it Nat. Mater.} {\bf 22} 570

\bibitem{KoskiJACS2012} Koski K J {\it et al} 2012  {\it J. Am. Chem. Soc.} {\bf 134} 13773 
\bibitem{ChaNanoLett2013} Cha J J {\it et al} 2013  {\it Nano Lett.} {\bf 13} 5913
\bibitem{YaoNatCommun2014} Yao J {\it et al} 2014  {\it Nat. Commun.} {\bf 5} 5670

\bibitem{WrayNatPhys2010} Wray L A {\it et al} 2010 {\it Nat. Phys.} {\bf 6} 855
\bibitem{TanakaPRB2012} Tanaka Y {\it et al} 2012 {\it Phys. Rev. B} {\bf 85} 12511
\bibitem{ChangScience2013} Chang C -Z {\it et al} 2013 {\it Science} {\bf 340} 167
\bibitem{Otrokov2DMater2017} Otrokov M M {\it et al} 2017 {\it 2D Mater.} {\bf 4} 025082
\bibitem{OtrokovJEPTL2017} Otrokov M M {\it et al} 2017, {\it JETP Lett.} {\bf 105} 297
\bibitem{ZhangNatPhys2022} Zhang H {\it et al} 2022 {\it Nat. Phys.} {\bf 18} 1425

\bibitem{ZhangNatCommun2018} Zhang J {\it et al} 2018 {\it Nat. Commun.} {\bf 9} 5289

\bibitem{CruzadoACSAEM2021} Cruzado H N {\it et al} 2021 {\it ACS Appl. Electron. Mater.} {\bf 3} 1071 
\bibitem{HsuPNAS2008} Hsu F-C {\it et al} 2008 {\it Proc. Natl. Acad. Sci. USA} {\bf 105} 14262
\bibitem{GuoPRB2010} Guo J {\it et al} 2010 {\it Phys. Rev. B} {\bf 82} 180520(R)
\bibitem{WangPRB2011} Wang A F {\it et al} 2011 {\it Phys. Rev. B} {\bf 83} 060512(R)
\bibitem{FangEPL2011} Fang M-H {\it et al} 2011 {\it Europhys. Lett.} {\bf 94} 27009
\bibitem{YingSciRep2012} Ying T P {\it et al} 2012 {\it Sci. Rep.} {\bf 2} 426
\bibitem{ScheidtEPJB2012} Scheidt E-W {\it et al} 2012 {\it Eur. Phys. J.} B {\bf 85} 279
\bibitem{KrztonJPCM2012} Krzton-Maziopa A {\it et al} 2012 {\it J. Phys.: Condens. Matter.} {\bf 24} 382202
\bibitem{BurrardNatMater2013} Burrard-Lucas M 2013 {\it Nat. Mater.} {\bf 12} 15
\bibitem{NojiPhysC2014} Noji T 2014 {\it Physica C} {\bf 504} 8
\bibitem{GiaoJPSJ2017} Phan G N {\it et al} 2017 {\it J. Phys. Soc. Jpn.} {\bf 86} 033706
\bibitem{MaletzPRB2014} Maletz J {\it et al} 2014 {\it Phys. Rev. B} {\bf 89} 220506(R)
\bibitem{NakayamaPRL2014} Nakayama K {\it et al} 2014 {\it Phys. Rev. Lett.} {\bf 113} 237001
\bibitem{TanNatMater2013} Tan S {\it et al} 2013 {\it Nat. Mater.} {\bf 12} 634
\bibitem{GiaoPRB2017} Giao P {\it et al} 2017 {\it Phys. Rev. B} {\bf 95} 224507

\bibitem{RiedlPRB2007} Riedl C {\it et al} 2007 {\it Phys. Rev. B} {\bf 76} 245406
\bibitem{MattauschPRL2007} Mattausch A and Pankratov O 2007 {\it Phys. Rev. Lett.} {\bf 99} 076802 
\bibitem{NakanoNanoLett2019} Nakano M {\it et al} 2019 {\it Nano Lett.} {\bf 19} 8806
\bibitem{LuNatMater2014} Lu X F {\it et al} 2014 {\it Nat. Mater.} {\bf 14} 325
\bibitem{BosPRB2007} Bos J W G {\it et al} 2007 {\it Phys. Rev. B} {\bf 75} 195203
\bibitem{NabokPRM2022} Nabok D {\it et al} 2022 {\it Phys. Rev. Mater.} {\bf 6} 034204
\bibitem{YehEPL2008} Yeh K-W {\it et al} 2008 {\it Europhys. Lett.} {\bf 84} 37002
\bibitem{FangPRB2008} Fang M H {\it et al} 2008 {\it Phys. Rev. B} {\bf 78} 224503 
\bibitem{MizuguchiJPSJ2009} Mizuguchi Y {\it et al} 2009 {\it J. Phys. Soc. Jpn.} {\bf 78} 074712
\bibitem{ChenPNAS2014} Chen T-K {\it et al} 2013 {\it Proc. Natl. Acad. Sci. USA} {\bf 111} 63
\bibitem{BaoCPL2011} Bao W {\it et al} 2011 {\it Chin. Phys. Lett.} {\bf 28} 086104
\bibitem{YePRL2011} Ye F {\it et al} 2011 {\it Phys. Rev. Lett.} {\bf 107} 137003
\bibitem{WeiNatPhys2012} Wei L {\it et al} 2012 {\it Nat. Phys.} {\bf 8} 126
\bibitem{YuPRM2020} Yu X-Q {\it et al} 2020 {\it Phys. Rev. Mater.} {\bf 4} 051402(R)

\bibitem{MiyataNatMater2015} Miyata Y {\it et al} 2015 {\it Nat. Mater.} {\bf 14} 775
\bibitem{ZhangAdvMater2023} Zhang W {\it et al} 2023 {\it Adv. Mater.} {\bf 35} 2209931
\bibitem{MazinPML2010} Mazin I I and Balatsky A V 2010 {\it Philos Mag. Lett.} {\bf 90} 731
\bibitem{JishiASTP2011} Jishi R A {\it et al} 2011 {\it Adv. Studies Theor. Phys.} {\bf 5} 703
\bibitem{PosternakPRL1983} Posternak M {\it et al} 1983 {\it Phys. Rev. Lett.} {\bf 50} 761
\bibitem{OhnoJPSJ1979} Ohno T {\it et al} 1979 {\it J. Phys. Soc. Jpn.} {\bf 47} 1125
\bibitem{IchinokuraACSNano2016} Ichinokura S {\it et al} 2016 {\it ACS Nano} {\bf 10} 2761
\bibitem{WangCPL2012} Wang Q Y {\it et al} 2012 {\it Chin. Phys. Lett.} {\bf 29} 037402
\bibitem{HeNatMater2013} He S {\it et al} 2013 {\it Nat. Mater.} {\bf 12} 605
\bibitem{ShiNatCommun2016} Shi X {\it et al} 2016 {\it Nat. Commun.} {\bf 8} 14988
\bibitem{LeeNature2014} Lee J J {\it et al} 2014 {\it Nature} {\bf 515} 245
\bibitem{RebecPRL2017} Rebec S N {\it et al} 2017 {\it Phys. Rev. Lett.} {\bf 118} 067002
\bibitem{SongNatCommun2019} Song Q {\it et al} 2019 {\it Nat. Commun.} {\bf 10} 758
\bibitem{LaiJACS2015} Lai X {\it et al} 2015 {\it J. Am. Chem. Soc.} {\bf 137} 10148
\bibitem{ChenNatCommun2015} Chen P {\it et al} 2015 {\it Nat. Commun.} {\bf 6} 8943
\bibitem{SugawaraACSNano2016} Sugawara K {\it et al} 2016 {\it ACS Nano} {\bf 10} 1341
\bibitem{WangPRB2019} Wang Y {\it et al} 2019 {\it Phys. Rev. B} {\bf 100} 241404
\bibitem{HajraACSNano2020} Hajra D {\it et al} 2020 {\it ACS Nano} {\bf 14} 15626
\bibitem{VaneyNatCommun2022} Vaney J B {\it et al} 2022 {\it Nat. Commun.} {\bf 13} 1462
\bibitem{MairoserNatCommun2015} Mairoser T {\it et al} 2015 {\it Nat Commun.} {\bf 6} 7716
\bibitem{KhareAdvMater2017} Khare A {\it et al} 2017 {\it Adv. Mater.} {\bf 29} 1606566
\bibitem{YangCIE2020} Yang S {\it et al} 2020 {\it Chem. Int. Ed.} {\bf 59} 465
\bibitem{WangAdvMat2022} Wang W {\it et al} 2022 {\it Adv. Mater.} {\bf 34} 2203220
\end{thebibliography}
\end{document}